\documentclass[twocolumn,english,prb,superscriptaddress]{revtex4}
\usepackage[T1]{fontenc}
\usepackage[latin9]{inputenc}
\usepackage{babel}
\usepackage{amsmath}
\usepackage{amssymb}
\usepackage{graphicx}
\usepackage{esint}
\usepackage[unicode=true,pdfusetitle,
 bookmarks=true,bookmarksnumbered=false,bookmarksopen=false,
 breaklinks=false,pdfborder={0 0 1},backref=section,colorlinks=false]
 {hyperref}

\makeatletter
\@ifundefined{textcolor}{}
{%
 \definecolor{BLACK}{gray}{0}
 \definecolor{WHITE}{gray}{1}
 \definecolor{RED}{rgb}{1,0,0}
 \definecolor{GREEN}{rgb}{0,1,0}
 \definecolor{BLUE}{rgb}{0,0,1}
 \definecolor{CYAN}{cmyk}{1,0,0,0}
 \definecolor{MAGENTA}{cmyk}{0,1,0,0}
 \definecolor{YELLOW}{cmyk}{0,0,1,0}
 }

\@ifundefined{definecolor}{\usepackage{color}}{}

\usepackage{times}\usepackage{mathrsfs}\usepackage{paralist}\usepackage{ifthen}\usepackage{dsfont}\@ifundefined{definecolor}{\usepackage{color}}{}

\def\be{\begin{equation}}
\def\ee{\end{equation}}
\def\bea{\begin{eqnarray}}
\def\eea{\end{eqnarray}}

\makeatother

\begin{document}

\title{Electronic Liquid Crystalline Phases in a Spin-Orbit Coupled Two-Dimensional
Electron Gas}

\author{Erez Berg}

\affiliation{Department of Physics, Harvard University, Cambridge, Massachusetts
02138, USA}

\author{Mark S. Rudner}

\affiliation{Department of Physics, Harvard University, Cambridge, Massachusetts
02138, USA}

\affiliation{IQOQI and Institute for Theoretical Physics,
University of Innsbruck, 6020 Innsbruck, Austria}

\author{Steven A. Kivelson}

\affiliation{Department of Physics, Stanford University, Stanford, California
94305, USA}

\date{\today }
\begin{abstract}
We argue that the ground state of a two-dimensional electron gas
with Rashba spin-orbit coupling realizes one of several possible
liquid crystalline or Wigner crystalline
phases in the low-density %, low-temperature
limit, even for
short-range repulsive electron-electron interactions (which decay
with distance with a power larger than 2). Depending on specifics
of the interactions, preferred ground-states
include an anisotropic Wigner crystal with an %extremely
increasingly anisotropic unit cell as the density decreases, a striped
or electron smectic phase, and a ferromagnetic phase which strongly
breaks the lattice point-group symmetry, \textit{i.e.}~exhibits nematic
order. %In contrast, with long range interactions, the usual hexagonal Wigner crystal is recovered. We discuss the magnetic properties of the ground state in either case.
 Melting of the anisotropic Wigner crystal or the smectic phase by
thermal or quantum fluctuations %SAKmight
gives rise to a non-magnetic nematic phase which preserves
time-reversal symmetry.
\end{abstract}
\maketitle

\section{introduction}

Enhancing the role of electron-electron interactions relative to that
of the kinetic energy often leads to interesting many-body effects.
Electron crystallization is an extreme example of this phenomenon.
%EB provides a rich setting in which to explore
%intriguing manifestations of
%the consequences of the quantum mechanical wavelike nature of electrons.
%At high densities, where the inter-particle spacing is small,
%classical arguments would predict that electrons should form an
%ordered crystalline structure to minimize their large repulsive
%Coulomb interaction energy. However, in this case the quantum
%mechanical kinetic energy costs associated with creating the short
%wavelength density modulations
%of the %high-density
%crystalline state %needed to form a high-density crystal
%overwhelm the potential energy benefits, leading to an equilibrium
%state with \textit{uniform density}, even down to zero temperature.
%In contrast,
At low densities, where electrons are far apart and the kinetic energy
cost of localization is low, Coulomb interactions dominate and the
electrons form an ordered state, known as a Wigner crystal\cite{Wigner1934}.

%Early on in the study of the interacting electron gas, it was realized
% that in the low-density limit, Coulomb interactions dominate over
% the kinetic energy, so the electrons form a crystalline state, known
% as the Wigner crystal\cite{Wigner1934}.
%Recently,
The nature of the crystallized electronic state has been intensely
investigated through Quantum Monte Carlo (QMC) calculations \cite{Tanatar1989,Drummond2009}.
These calculations support the existence of the crystalline phase,
though the density at which they find crystallization is typically
significantly lower than what heuristic arguments would suggest. The
Wigner crystal has also been sought experimentally. % in
%Despite the difficulty of experiments in the ultra-low density regime where crystallization is expected, there
% and in experiments on
Despite difficulties associated with reaching the ultra-low density
regime where crystallization is expected, evidence that the Wigner
crystal phase may have been realized has been reported for experiments
on ever-cleaner samples of two-dimensional electron gases (2DEGs)
in semiconductor quantum wells\cite{Yoon1999}. %Although there is some
\begin{figure}[b]
\includegraphics[width=0.5\textwidth]{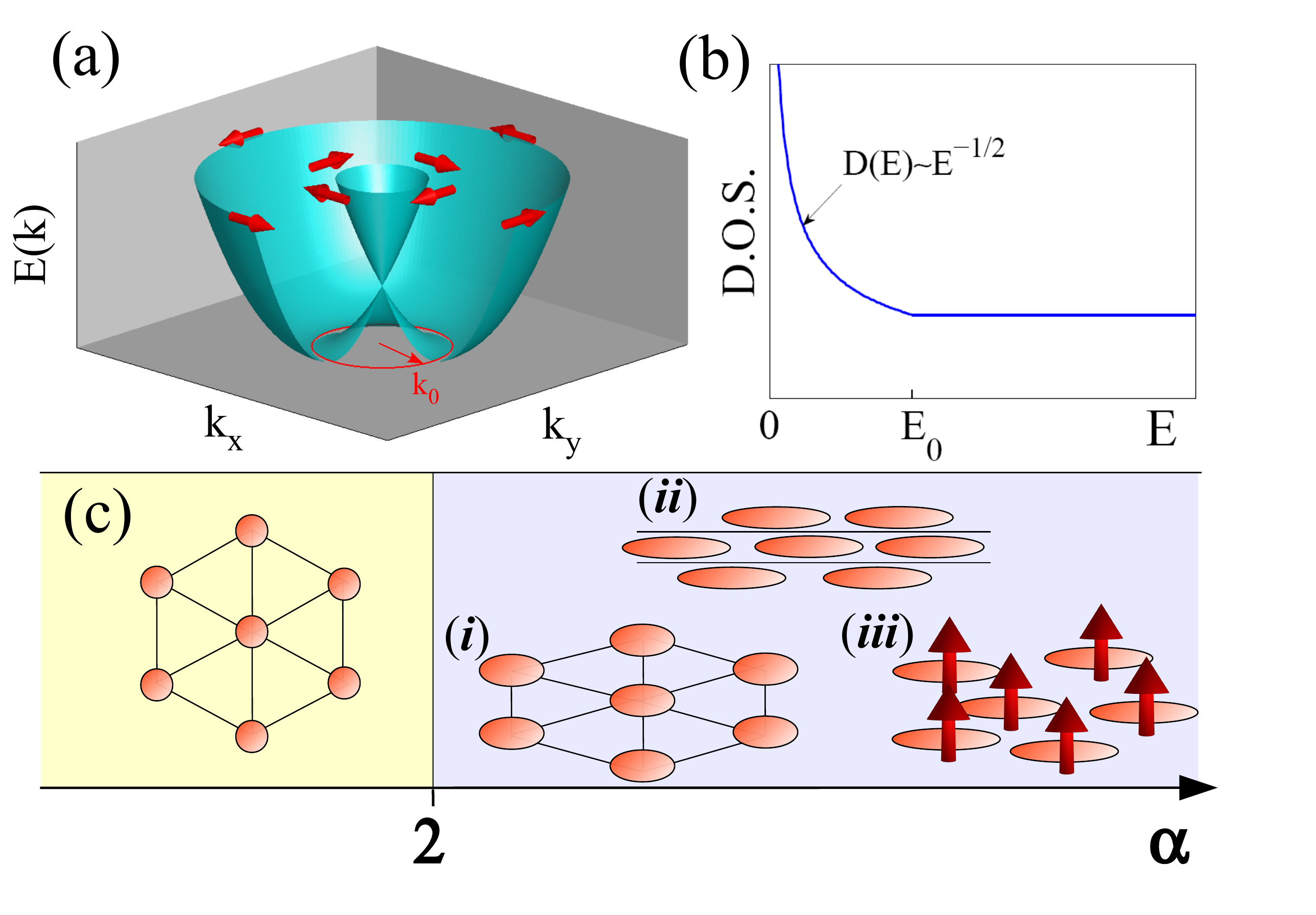}
\caption{(a) Dispersion of a particle with Rashba SOC. The minimum
of the dispersion occurs on a ring in $k$ space, marked by a red
circle. The red arrows show the spin polarization of the different
Bloch states. (b) Density of states as a function of energy,
corresponding to the dispersion shown in (a). Near the band
bottom, the density of states diverges as $\rho(E)\sim(E)^{-1/2}$.
(c) Schematic phase diagram in the low-density limit with
repulsive
electron-electron interactions % as a function of the exponent $\alpha$ with which the interactions
which decay at long distances as $V\sim r^{-\alpha}$. For $\alpha\le2$,
the ground state is an isotropic Wigner crystal %is {}``isotropic,''
(\textit{i.e.} it preserves a discrete rotational symmetry,
C$_{n}$ with $n>2$). For $\alpha>2$, states with a further broken
rotational symmetry are favored. (i) represents an anisotropic
Wigner crystal with a unit cell which becomes parametrically
anisotropic in the low--density limit. (ii) and (iii) represent
snapshots of a smectic state and a
ferromagnetic nematic liquid state, respectively.%, a smectic state (which
%breaks translational invariance in one direction) and a
%non-magnetic nematic state.
} \label{fig:density}
\end{figure}

%%%%%%%%%%%%%%%%%%%%%%%%%%%%%%%%%%%%%%%%%%%%%%%%%%%%%%%%%%%%%%%%%%%%%%%%%%%%%%
In this paper, we study the low density limit of the 2DEG system with
Rashba spin-orbit coupling (SOC), which is present whenever the 2DEG
lacks inversion symmetry\cite{Rashba1960}. This is the case, for
example, when the 2DEG is confined in an asymmetric quantum well,
or if it is formed at the surface of a three-dimensional material.
As shown in Fig.\ref{fig:density}a, the resulting dispersion relation
has an extended (highly degenerate) minimum which forms a ring in
momentum space. The low-energy density of states exhibits a divergent
van Hove singularity, $\rho(\varepsilon)\sim\varepsilon^{-1/2}$,
akin to the behavior of a one-dimensional system (see Fig.~\ref{fig:density}b).
This is in striking contrast to the usual behavior $\rho(\varepsilon)\sim{\rm const}$,
familiar for two-dimensional systems without spin-orbit coupling.
Thus the Rashba SOC %helps to enhance
greatly enhances the role of interactions relative to that of kinetic
energy in the low-density limit. %Consequently, the role of interactions is enhanced by Rashba SOC in the low-density limit.

For a system with Coulomb interactions, $V(r)\sim1/r$, the ground
state at low densities is a Wigner crystal, just as for a 2DEG without
spin orbit coupling. Remarkably, however, we find that %In the low-density limit, the Coulomb interacting system must form
%a Wigner crystal, just like the regular 2DEG. However,
with Rashba SOC, %a Wigner crystal appears to be favored over the uniform
broken symmetry states appear to be favored over the uniform Fermi
liquid (UFL) state even for \textit{short-range} %electron-electron
interactions, %(e.g.~if the Coulomb interactions are screened by a metallic gate), which we model by a power-law dependence
 $V(r)\sim1/r^{\alpha}$ with $\alpha>2$. (In particular, %for the Coulomb interactions screened by a metallic gate, $\alpha=3$.)
note that the Coulomb interaction screened by a metallic gate is
described by $\alpha = 3$.) The instability in this case occurs at
an electron density $n$ for which
the Fermi energy %becomes
is smaller than an energy scale set by the SOC. %Moreover, we find
%that if the electron-electron interactions fall off as a power law
%with distance with a power greater than 2, the Wigner crystal has
%an anisotropic unit cell, and the anisotropy becomes
%parametrically large in the low-density limit.

We have investigated candidate ordered states by constructing
variational wave-functions, determining the patterns of broken
symmetry which minimize their variational energies, and then
comparing the energy to that of the UFL. We consider the following
types of broken symmetry states:
%Explicit broken symmetry states we have investigated include:
1) Wigner crystalline (WC) states, \textit{i.e.}~insulating states
%with a pattern of translation symmetry breaking
with only discrete translational symmetry corresponding to one
electron per unit cell, allowing for %and
various possible crystal structures corresponding to different
patterns of rotation symmetry breaking; 2) an electron smectic
state which breaks translational symmetry in only one direction,
and can be viewed as a partially melted version of an anisotropic
WC; 3) a ferromagnetic nematic state which preserves translation
symmetry, but breaks time reversal symmetry and rotational
symmetry - this state is invariant under time reversal followed by
a rotation by $\pi$ around the symmetry axis. Note that we refer
to a WC as anisotropic when only a discrete 2-fold rotation
symmetry (C$_{2}$) remains unbroken. In the limit of low density,
each of these ordered states has parametrically lower energy (in
powers of the density $n$) than the UFL. This strongly suggests
that the UFL is unstable at low densities. In contrast, the energy
balance between different broken symmetry phases is more delicate,
and may well depend on long-distance fluctuational effects that
are not well captured by variational wavefunctions; we will return
to this point in the final section of the paper.

The nature of the low-density instability of the UFL, and the origin
of the strong tendency of the system to a nematic pattern of rotation
symmetry breaking (whether or not it is accompanied by other patterns
of symmetry breaking) can be most easily seen by studying candidate
WC wavefunctions. A schematic version of the resulting WC phase diagram
%of a 2DEG in the low-density limit
 is shown in Fig. \ref{fig:density}c as a function of %$\alpha$,
the exponent $\alpha$. %with which the electron-electron interactions fall off at large distances.
 For $\alpha>2$ (short-range interactions), the unit cell of the
Wigner crystal becomes increasingly anisotropic, %SAKdiverging
with an aspect ratio that diverges in the low-density limit. This
unusual behavior can be traced back to the form of the single-particle
dispersion, Fig.\ref{fig:density}a, which is strongly anisotropic
in the directions perpendicular and tangential to the ring-like minimum.
%in the presence of Rashba SOC (Fig. \ref{fig:density}a). The minimum
%of this dispersion occurs on a ring in momentum space, giving rise
%to a divergence of the density of states at low energies.
%***Move this discussion to the end?***We note that such behavior is not unique to a 2DEG with Rashba SOC, and can occur in other contexts such as paired-fermion systems\cite{Yang2006}. In addition, the arguments in support of an anomalously anisotropic Wigner crystal do not rely on the particle statistics. Therefore, we expect our conclusions to apply to bosonic systems with effective Rashba SOC (see e.g.~Refs.\onlinecite{Stanescu2008,Wang2010,Gopalakrishnan2011}) as well. ****

A complimentary view can be obtained by considering a ferromagnetic
state. Because of the Rashba SOC, the orientation of the magnetization
vector (which is always in-plane) necessarily defines a preferred
nematic axis - any ferromagnetic state must necessarily be nematic,
although a non-magnetic nematic phase is possible. For example, a
magnetic moment in the y direction implies a special role for the
point $\vec{k}_{0}$ in Fig.~1a, which defines the point on the ring
of minimum dispersion with the largest possible value of $k_x$. %component.
At low density, the resulting Fermi surface forms an ellipse encircling this
special point. As in the usual Stoner theory of ferromagnetism, %there is a short-distance gain in
spin polarization lowers the interaction energy via %due to
the Pauli-exclusion principle, which helps electrons to avoid each
other at short distances. However, the cost in kinetic energy is
parametrically smaller than in a conventional FL, owing to the
divergent density of states. The variational energy we find for
the ferromagnetic nematic state differs from that of the
anisotropic crystal only by a numerical constant for
$2<\alpha\leq3$, so it is not possible, on the basis of the
present considerations, to confidently determine which (if either)
is the preferred ground state. For $\alpha > 3$, the ferromagnetic
nematic state has parametrically lower energy than the anisotropic
WC, suggesting that it is a better candidate ground state. The
scaling of the ground state energy as a function of the electron
density for each type of state considered in this paper is listed
in Table \ref{tab:E}.

\begin{table}
\caption{Scaling of the ground state energy per particle as a function of the
electron density $n$ in the low density limit, for the various candidate
ground states considered in this paper: the nematic ferromagnetic
(FM) state, the anisotropic Wigner crystal (AWC), and the smectic.
For $\alpha<2$, the isotropic Wigner crystal always has the lowest
energy.\label{tab:E}}

\vspace{0.1in}
\begin{tabular*}{0.48\textwidth}{@{\extracolsep{\fill}} l c c c} \hline \hline state & $2<\alpha\le3$ & $3<\alpha\le4$ & $4<\alpha$ \tabularnewline \hline nematic FM   & $n^{2\left(1-\frac{1}{\alpha}\right)}$ & $n^{2\left(1-\frac{1}{\alpha}\right)}$ & $n^{\frac{3}{2}}$ \tabularnewline \hline AWC or smectic & $n^{2\left(1-\frac{1}{\alpha}\right)}$ & $n^{\frac{4}{3}}$ & $n^{\frac{4}{3}}$  \tabularnewline \hline \hline \end{tabular*}
\end{table}

This paper is organized as follows. The model is described in Sec.~\ref{sec:model}. In Sec.~\ref{sec:WC}, we explain the basic physics
leading to anisotropic Wigner crystal formation. % in the presence of Rashba SOC.
We analyze three cases: contact interactions, extended short-range
interactions (which fall off with distance with a power which is
larger than 2), and long-range interactions. Considerations
related to the magnetic structure of the Wigner crystal phase are
discussed in Sec.~\ref{sec:magnetic}. In
Sec.~\ref{sec:Smectic-state} we discuss melting the Wigner crystal
partially to obtain a smectic state. In
Sec.~\ref{sec:Ferromagnetic-nematic-state} we discuss the
ferromagnetic nematic state. Sec.~\ref{sec:Exp} presents a
proposed schematic phase phase diagram for a 2DEG with SOC and
screened Coulomb interactions. In Sec.~\ref{sec:real} we discuss
possible realizations in electronic and atomic systems. Appendix
\ref{sec:App} presents the solution of a single Rashba particle in
a box problem, which is crucial for the arguments regarding the
anisotropic Wigner crystal and the smectic phases, and Appendix
\ref{App:HF} presents the details of the Hartree-Fock analysis of
the ferromagnetic nematic state.

\section{Model}

\label{sec:model}

We consider a 2DEG with Rashba SOC and repulsive electron-electron
interactions, described by the Hamiltonian (in units with $\hbar=1$)
\begin{eqnarray}
H&=&\sum_{j}\left\{
\frac{1}{2m}\left[-\nabla_{j}^{2}-\frac{2k_{0}}{i}(\nabla_{j}\times{\hat{z}})\cdot\vec{\sigma}_{j}\right]+E_0
\right\} \notag
\\ &+&\frac{1}{2}\sum_{l\neq
j}V(|\vec{r}_{l}-\vec{r}_{j}|)\text{.}\label{H}
\end{eqnarray}
 Here, $m$ is the electronic effective mass, $k_{0}$ is a parameter
that characterizes the strength of the SOC, $E_0\equiv
k_0^2/(2m)$, $\vec{\sigma}_{j}$ is the vector of Pauli matrices
which act on the spin of electron $j$, and $V(|\vec{r}|)$ is the
(repulsive) electron--electron interaction
potential. %We work in units in which $\hbar=1$.
The Hamiltonian (\ref{H}) is invariant under translations, under
rotations around the $z$ axis, and under mirror reflections about
the $x$ and $y$ axes, $M_{x}$ and $M_{y}$, but \emph{not} under
inversion $\vec{r}\rightarrow-\vec{r}$. %Rashba SOC\ arises whenever the 2DEG arises in situations in which the system lacks inversion symmetry.
% This can happen, for instance, when the 2DEG\ is confined in an asymmetric quantum well, or if it is confined to a surface of a three-dimensional material.

Below we consider cases in which, at large inter-particle separation
$r$, the interaction potential %$V\left(\vec{r}\right)$
decays as a power law: % with distance:
\begin{equation}
V\left(|\vec{r}|\right)\sim\frac{V_{0}}{r^{\alpha}}\text{.}\label{alpha}
\end{equation}
 We distinguish between long-range and short-range interactions, which
are characterized by $\alpha\leq2$ and $\alpha>2$, respectively.
The bare Coulomb interaction is described by $\alpha=1$, while screening
can lead to $\alpha>1$. In particular, screening due to a nearby
metallic gate top gate leads to $\alpha=3$. %This case also can be used to describe the behavior of interacting dipolar atoms or molecules in an optical

In the absence of interactions, $V(|\vec{r}|)=0$, Eq.~(\ref{H})
yields the single-particle dispersion law (see
Fig.~\ref{fig:density}a):
%associated with the Hamhas the form
\begin{equation}
E(\vec{k})=\frac{1}{2m}\left(k^{2}\pm2k_{0}k\right)+E_0,\label{Ek}
\end{equation}
 where $\vec{k}$ is the electron momentum ($\hbar=1$), and $k=|\vec{k}|$.
The minimum kinetic energy occurs for any value of momentum
falling on a ring of radius $k=k_{0}$, with a minimal value of
$E=0$. The corresponding density of states, shown in
Fig.~\ref{fig:density}b,
is given by %as a function of energy is
\begin{equation}
D\left(E\right)=\left\{ \begin{array}{ccl}
\frac{m}{\pi}\sqrt{\frac{\left\vert E_{0}\right\vert }{E}} &  & (E<E_0),\\
\frac{m}{\pi} &  & (E>E_0).
\end{array}\right.\label{DOS}
\end{equation}
 For $E>E_0$, the density of states is independent of energy, just
as for a usual 2DEG without spin-orbit coupling. However, for
$E\rightarrow 0$, the density of states diverges as $1/\sqrt{E}$.
Because this divergence will play a crucial role in the analysis
below, we comment briefly on its origin. The divergence comes from
the fact that the minimum of kinetic energy is infinitely
degenerate, occurring everywhere on a ring in momentum space,
rather than at a single point or finite set of points. In the
presence of crystalline anisotropy
(which %SAKis present
 manifests itself through corrections to the effective mass approximation
in real materials), %If we introduce crystalline anisotropy through crystal field terms (which are inevitably present in any real material),
the divergence is cut off near the band bottom and the density of
states goes to a constant. However, as long as $k_{0}a/\pi\ll1$,
where $a$ is the lattice constant (\emph{i.e.}~as long as the
spin-orbit coupling is weak), the crystal field anisotropy terms
are small and Eq.~(\ref{DOS}) provides a good approximation down
to energies of order $|E_{0}|(k_{0}a)^{2}$ above the band bottom.

\section{Wigner crystal}

\label{sec:WC}

\subsection{Instability of the Fermi liquid state}

We begin by considering the stability of the uniform (Fermi liquid)
state. According to Eq.~(\ref{DOS}), at low densities, the Fermi
energy $\varepsilon_{F}$ %(measured from the band bottom $E_{\min}$)
is
\begin{equation}
\varepsilon_{F}=\frac{\pi^{2}n^{2}}{4m^{2}\left|E_{0}\right|}\text{,}
\end{equation}
 where $n$ is the density of electrons per unit area. Thus, the kinetic
energy per particle in the homogeneous state is
\begin{equation}
\bar{\varepsilon}_{\text{kin}}=\frac{1}{n}\int_{0}^{\varepsilon_{F}}\varepsilon
D\left(\varepsilon\right)d\varepsilon=\frac{\pi^{2}n^{2}}{12m^{2}\left|E_{0}\right|}\text{.}
\end{equation}
 The potential energy per particle, on the other hand, is
\begin{equation}
\bar{\varepsilon}_{\text{pot}}=\frac{1}{2n\Omega}\int d\vec{r}d\vec{r}^{\prime}V\left(\vec{r}-\vec{r}'\right)\langle n(\vec{r})n(\vec{r}')\rangle\text{,}\label{epot}
\end{equation}
 where $\Omega$ is the total area of the system, $n(\vec{r})$ is
the local density at position $\vec{r}$, and $\langle\cdots\rangle$
denotes averaging in the uniform (Fermi gas) state. In the low--density
limit, for short-range interactions ($\alpha>2$ in Eq. \ref{alpha}),
$\bar{\varepsilon}_{\text{pot}}\propto n$. For long-range interactions,
the right hand side of Eq.~(\ref{epot}) diverges. %one needs to add
Here a neutralizing background must be taken into account, leading
to $\bar{\varepsilon}_{\text{pot}}\propto n^{\alpha/2}$. We see that
in all cases, in the low-density limit, $\bar{\varepsilon}_{\text{pot}}\gg\bar{\varepsilon}_{\text{kin}}$,
suggesting that the uniform state is unstable to forming some sort
of order. One possibility is that at asymptotically low densities,
the ground state is a Wigner crystal, as in a 2DEG with no SOC. Note, however, that here,
%however, that
in the presence of Rashba SOC, this instability occurs \emph{even
for short-range interactions}.

\subsection{Contact interactions}

\label{sec:contact}

For simplicity, we begin by considering the case of the shortest range
interactions: %\textit{i.e.}
repulsive {}``contact'' interactions. We start with a variational
wavefunction which minimizes the interaction energy, taking each electron
to be confined to a rectangular box of dimensions $L_{x}\times L_{y}$,
with different boxes non-overlapping. This is a zero energy eigenstate
of the potential energy operator. %EB, and hence a (non unique) groundstate.
%and the potential energy is exactly zero. The total energy is equal to the kinetic energy.
 In order for the boxes to tile the plane, $L_{x}$ and $L_{y}$ are
constrained by the condition
\begin{equation}
L_{x}L_{y}=\frac{1}{n}.
\end{equation}
 %Now, we are faced with the problem of optimizing the aspect ratio
%$\ell =L_{x}/L_{y}$, with $n$ fixed.
Thus, we have a single variational parameter, the aspect ratio $\eta\equiv L_{x}/L_{y}$,
which we use to minimize the kinetic energy of the trial state. The
kinetic energy per particle in the variational state is given by the
ground state energy of a single particle in a box with Rashba SOC.
This problem is investigated, both analytically and numerically, in
Appendix A. Surprisingly, unlike the case with no SOC, the ground
state energy in the low--density limit is minimal for $\eta\neq1$.
%EBIn the low--density limit, the minimum occurs
%for $\ell \gg 1$.
In the $\eta\gg1$ limit, we find the following expression for the
ground state energy as a function of $\eta$ and $n$:\cite{comment-ell}
\begin{equation}
\varepsilon\left(n,\eta\right)=\frac{n}{2m}\left(A\eta^{-1}+\frac{Bn}{k_{0}^{2}}\eta^{2}\right),\label{Enl}
\end{equation}
 where $A$ and $B$ are numbers of order unity, see Eq.~(\ref{eLxLy})
in Appendix A. Minimizing Eq.~(\ref{Enl}) with respect to $\eta$,
we find that the optimal aspect ratio $\eta^{\star}$ scales as
\begin{equation}
\eta^{\star}\sim\left(n/k_{0}^{2}\right)^{-\frac{1}{3}},\label{Asp}
\end{equation}
 and the ground state energy per particle %(measured relative to $E_{\min}$)
scales as
\begin{equation}
\varepsilon^{\star}\left(\eta^{\star}\right)\sim\left|E_{0}\right|\left(n/k_{0}^{2}\right)^{\frac{4}{3}}\text{.}\label{Ekin}
\end{equation}
 Therefore, in the low--density limit, we get that the energy per
particle of this anisotropic Wigner crystal state is parametrically
lower than %SAKthe energy
that of the uniform state, which scales as $n$. Note that,
consistent with our assumptions, the optimal aspect ratio of the
unit cell in the Wigner crystal becomes parametrically large at
low densities.

%Heuristically,
The fact that the kinetic energy is minimal for an anisotropic box
can be understood as follows. Suppose that the ground state wavefunction
is a superposition of plane waves with wavevectors close to some wavevector
$\vec{k}^{\star}$ of length $k_{0}$, for which the dispersion (\ref{Ek})
is minimal. Near $\vec{k}^{\star}$, the dispersion is quadratic in
the radial direction, while it is anomalously flat (quartic) in the
transverse direction. Therefore confinement in the direction perpendicular
to $\vec{k}^{\star}$ is less costly than confinement parallel to
$\vec{k}^{\star}$, and the optimal aspect ratio is such that the
box is long in the direction of $\vec{k}^{\star}$, and short in the
transverse direction.

\smallskip{}

\subsection{Extended short-range interactions\label{sub:Extended-short-range-interaction}}

We now turn to the case of extended, short--range interactions, which
corresponds to $2<\alpha<\infty$. We show that in this case, as in
the case of contact interactions, the Wigner crystal state is extremely
anisotropic in the low--density limit.

In the case of extended interactions, the potential energy in the
Wigner crystal phase cannot be neglected. To estimate the
potential energy, we consider the same variational wave function
as before, in which the particles are localized in an array of
non-overlapping
$L_{x}\times L_{y}$ boxes. %For simplicity in evaluating
To estimate the potential energy, we will replace the wavefunction
of each particle by a constant, such that the density is uniform,
$n=1/(L_{x}L_{y})$; the parametric dependence of the energy on $n$
and $\eta$ should not depend on this assumption. Let us focus on
the anisotropic limit, $L_{x}\gg L_{y}$, assuming that this is the
optimal configuration. The interaction energy of a given particle
with all the other particles is

\begin{eqnarray}
\bar{\varepsilon}_{\text{v}}\left(n,\eta\right) & \approx & 2(U_{1}+U_{2}),\label{Evv}
\end{eqnarray}
 where $U_{1}$ and $U_{2}$ are given by \begin{widetext}
\begin{align}
U_{1} & =\frac{1}{L_{x}^{2}L_{y}^{2}}\int_{0}^{L_{x}}dx\int_{0}^{L_{y}}dy\int_{-\infty}^{\infty}dx'\int_{L_{y}}^{\infty}dy'\frac{V_{0}}{\left[(x-x')^{2}+(y-y')^{2}\right]^{\alpha/2}}=V_{0}n^{\frac{\alpha}{2}}\eta^{\frac{\alpha}{2}-1}C_{1}
\end{align}
 and
\begin{align}
U_{2} & =\frac{1}{L_{x}^{2}L_{y}^{2}}\int_{0}^{L_{x}}dx\int_{0}^{L_{y}}dy\int_{L_{x}}^{\infty}dx'\int_{0}^{L_{y}}dy'\frac{V_{0}}{\left[(x-x')^{2}+(y-y')^{2}\right]^{\alpha/2}}=V_{0}n^{\frac{\alpha}{2}}\eta^{\frac{\alpha}{2}-2}\left[C_{2}+O\left(\eta^{2-\alpha}\right)\right].
\end{align}
 \end{widetext} $C_{1}$ and $C_{2}$ are dimensionless constants
which depend on $\alpha$. In the $\eta\gg1$ limit, we get that $U_{1}\gg U_{2}$,
and therefore we neglect the latter. Now, if we assume that $\eta\sim n^{-\frac{1}{3}}$,
as Eq.~(\ref{Asp}) suggests in the case of contact interactions,
we find
\begin{equation}
\bar{\varepsilon}_{\text{v}}\sim n^{\frac{1}{3}\left(1+\alpha\right)}\text{.}
\end{equation}
 We see that, for $\alpha>3$, $\bar{\varepsilon}_{\text{v}}$ becomes
negligible compared to the kinetic energy $\varepsilon^{\star}(\eta^{\star})\sim n^{4/3}$,
Eq.~(\ref{Ekin}), in the $n\rightarrow0$ limit. Therefore, Eqs.~(\ref{Asp})
and (\ref{Ekin}) are not modified in this case. For $\alpha<3$,
$\bar{\varepsilon}_{\text{v}}$ dominates in the low-density limit,
and we should consider both the kinetic and potential energies, Eqs.~(\ref{Ekin}),(\ref{Evv}):
\begin{eqnarray}
\bar{\varepsilon}_{\text{tot}} & = & \bar{\varepsilon}_{\text{v}}\left(n,\eta\right)+\varepsilon\left(n,\eta\right)\nonumber \\
 & \approx & \frac{n}{2m}\left(A\eta^{-1}+\frac{Bn}{k_{0}^{2}}\eta^{2}\right)+C_{1}V_{0}n^{\frac{\alpha}{2}}\eta^{\frac{\alpha-2}{2}}\text{.}\label{E}
\end{eqnarray}
 Minimizing with respect to $\eta$ and keeping only the most singular
term as $n\rightarrow0$ gives
\begin{equation}
\eta^{\star}\sim\frac{1}{\left(2mV_{0}\right)^{2/\alpha}}n^{\frac{2}{\alpha}-1}\text{ \ \ (}2<\alpha<3\text{)}.\label{ell1}
\end{equation}
 We see that, for $2<\alpha<3$, $\eta$ still becomes parametrically
large in the $n\rightarrow0$ limit. Inserting Eq.~(\ref{ell1})
back into Eq.~(\ref{E}), we get
\begin{equation}
\varepsilon_{\text{tot}}^{\star}\sim\frac{1}{m}\left(2mV_{0}\right)^{2/\alpha}n^{2\left(1-\frac{1}{\alpha}\right)}\text{ \ \ \ (}2<\alpha<3\text{)}.\label{etot}
\end{equation}
 % such that,
Thus for $2<\alpha<3$ the anisotropic Wigner crystal has parametrically
lower energy per particle than the uniform state. %SAK, whose energy scales as $n$.

\subsection{Long-range interactions}

For $\alpha<2$ (long range interactions), the Wigner crystal in
the low density limit has the same hexagonal ($C_{6}$) symmetric
triangular structure as the classical crystalline phase which
minimizes the potential energy. We will refer to the hexagonal
crystal as {}``isotropic'', as opposed to the {}``anisotropic''
crystal described previously, which has a lower symmetry. To show
that the Wigner crystal is isotropic for $\alpha<2$, we note that
the potential energy in a classical crystal scales as
\begin{equation}
\varepsilon_{\text{WC}}\sim n^{\frac{\alpha}{2}}\label{E1}.
\end{equation}
% relative to the uniform state.
If, in the Wigner crystal phase, each electron is confined to a
region whose dimension is some fraction of the mean inter-electron
distance, the kinetic energy cost of forming the crystal scales as
the density $n$, as in the case without
SOC\cite{comment-isotropic}. (Here, unlike before, we assume that
the region to which the electron is confined has an aspect ratio
of order unity.) Therefore, in the low-density limit,
crystallization yields a potential energy gain which overwhelms
the kinetic energy cost. Thus, to first approximation, we may
ignore the kinetic energy. The ground state is therefore a
hexagonal Wigner crystal, and is not qualitatively affected by the
Rashba SOC.

% By dimensional analysis, the shear
%modulus of the classical crystal is of the order of $\varepsilon_{\text{WC}}$.
%In the Wigner crystal, each particle is trapped in an approximately
%harmonic potential formed by the potential of its neighbors. Expanding
%this potential to second order in the displacement of the particle
%from the minimum, we get that the characteristic frequency $\omega$
%of the harmonic well satisfies
%\begin{equation}
%\omega^{2}\sim\frac{1}{r^{\alpha+2}}\sim n^{\frac{\alpha}{2}+1}
%\end{equation}
% where $r$ is the average inter-particle distance. Correspondingly,
%the characteristic size of the wavefunction is
%\begin{equation}
%r_{0}\sim\omega^{-\frac{1}{2}}\sim r^{\frac{\alpha}{4}+\frac{1}{2}}.\label{r0}
%\end{equation}
% Thus for $\alpha<2$ we see that the electrons are well-localized,
%$r_{0}\ll r$, ensuring the self-consistency of this approach.\cite{comment-harmonic}
%
%The zero-point energy (kinetic energy per particle) associated with
%localization in the crystallized state scales as
%\begin{equation}
%\varepsilon_{0}\sim\omega\sim n^{\frac{\alpha}{4}+\frac{1}{2}}.\label{E0}
%\end{equation}
% Comparing Eq. (\ref{E0}) to Eq. (\ref{E1}), we see that for $\alpha<2$
%the zero-point energy is negligible compared to the potential energy
%in the low-density limit. Since any possible energy gain from distortion
%is bounded from above by $\varepsilon_{0}$, we conclude that no distortion
%is expected in this case.

\section{Magnetic properties of the Wigner crystal}

\label{sec:magnetic}

So far, we have ignored the magnetic degrees of freedom. The ground
state of a Rashba particle in a box is two-fold degenerate, according
to Kramers' theorem. Correspondingly, the variational wavefunction
we considered (in which electrons occupy non-overlapping boxes) is
$2^{N}$-fold degenerate, where $N$ is the number of electrons. This
degeneracy is lifted by exchange interactions.

First, we elucidate the nature of the Kramers pair of ground states
of a single electron in a box. Because of the spin-orbit coupling,
these states are not eigenstates of the spin operator $\vec{\sigma}$.
However, in the anisotropic ($\eta=L_{x}/L_{y}\gg1$) limit, there
are particular linear combinations of the two ground states which
are approximately polarized in the $\pm\hat{y}$ directions, where
$\hat{y}$ is the narrow direction of the unit cell. The expectation
values of all other spin components are small; this is true for \emph{any}
choice of basis in the ground state Hilbert space.

\begin{figure}
\includegraphics[scale=0.5]{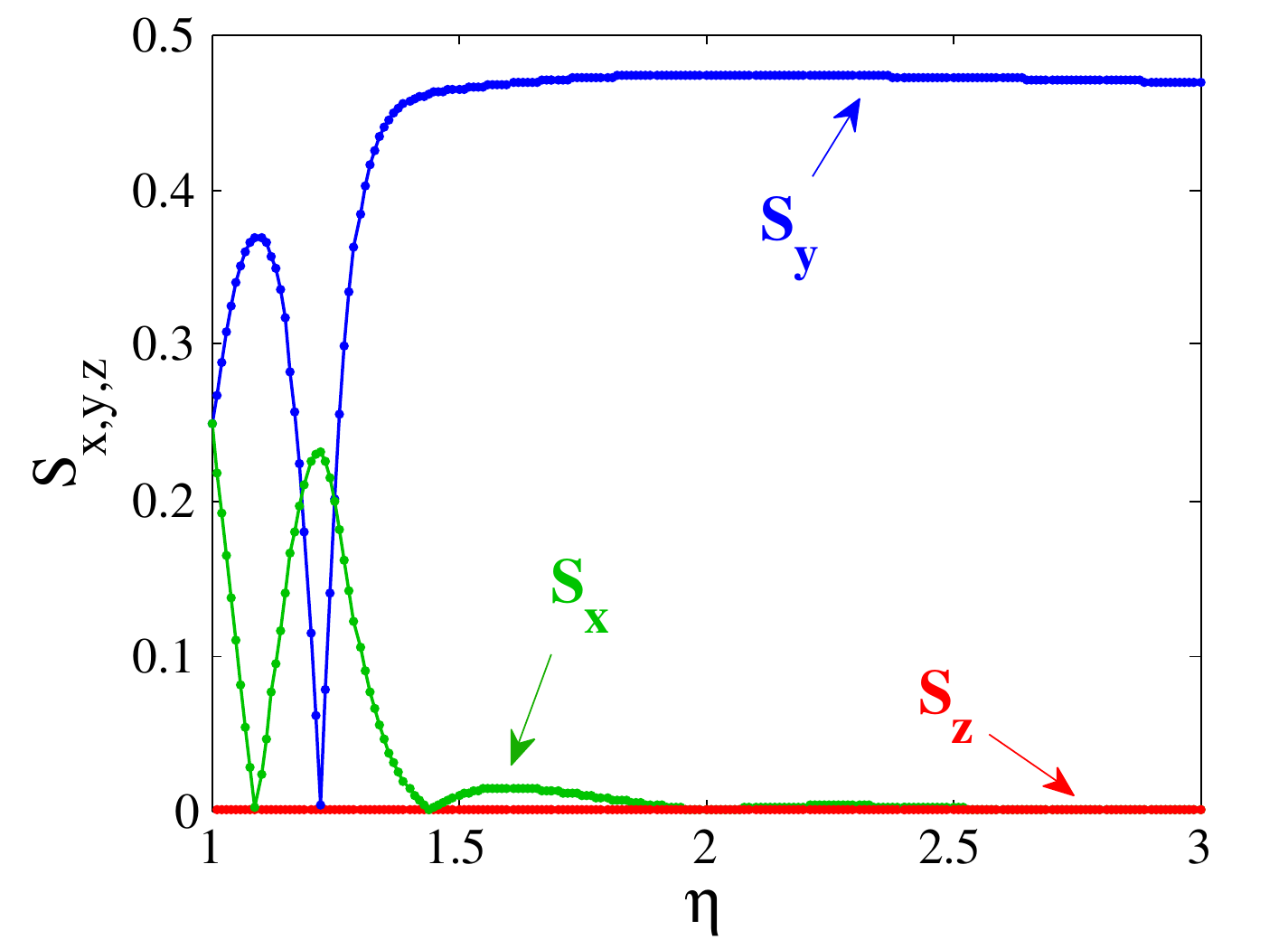} \caption{The quantities $S_{x,y,z}$, defined in Eq. \ref{eq:S} \textit{vs.}
the aspect ratio $\eta=L_{x}/L_{y}$ of a box with fixed area $\Omega=75/k_{0}^{2}$.
When $L_{x}/L_{y}=1$, $S_{x}=S_{y}$; as $L_{x}/L_{y}$ increases,
$S_{y}$ becomes close to $1$ and $S_{x}\rightarrow0$.}

\label{fig:S}
\end{figure}

To demonstrate this, we calculate the quantities
\begin{equation}
S_{i}\equiv\sqrt{\frac{1}{2}\sum_{\alpha,\beta=1,2}\left|\langle\alpha\vert s_{i}\vert\beta\rangle\right|^{2}},\label{eq:S}
\end{equation}
 where $\vert\alpha=1,2\rangle$ are the two ground states obtained
from the numerical solution of the particle in a box problem (see
Appendix \ref{sec:App}), and $s_{i}=\sigma_{i}/2$ where $\sigma_{i=x,y,z}$
are Pauli (spin) matrices. As defined, $S_{i}$ is the maximum expectation
value of the spin component $i$ in the ground state manifold spanned
by the Kramers pair $\vert\alpha=1,2\rangle$. The values of $S_{i=x,y,z}$
as functions of the aspect ratio $\eta$ are shown in Fig.~\ref{fig:S}.
As $\eta$ increases, $S_{y}$ becomes close to 1/2, while $S_{x,z}\rightarrow0$.
This can be understood as a consequence of the fact that when $L_{x}\gg L_{y}$,
the ground states contain mostly components with momenta close to
$\vec{k}=\pm k_{0}\hat{x}$, %whose spin is polarized
with spin polarizations close to the $\pm\hat{y}$ directions.

%%%%%%%%%%%%%%%%%%%%%%%%%%%%%%%%%%%%%%%%%%%%%%%%%%%%%%%%%%

%%%%%%%%%%%%%%%%%%%%%%%%%%%%%%%%%%%%%%%%%%%%%%%%%%%%%%%%%%

The magnetic degrees of freedom in the anisotropic Wigner crystal
can therefore be thought of as Ising-like spins, which are polarized
in the $\pm\hat{y}$ directions. These spins are coupled by exchange
processes, which generate N-body interactions\cite{Roger1984,Voelker2001,Bernu2001}.
In addition, because of the spin-orbit coupling, Van der Waals-like
interactions generate spin-spin terms\cite{Gangadharaiah2008}, which
are not exponentially suppressed in the Wigner crystal phase. A detailed
estimate of these interactions is complicated, and we defer their
analysis for later work.

%However, we note that the magnetic exchange interactions are likely
%to be frustrated.\mpar{make connection to FM?} Therefore, the crystal phase should have a large
%low-temperature entropy, much like the ordinary Wigner crystal\cite{Spivak2004}.
%In addition, it should be relatively easy to align the magnetic degrees
%of freedom using an in-plane field. Since the magnetic response of
%the Wigner crystal is strongly anisotropic, an in-plane magnetic field
%tends to align the crystal with either the short or the long direction
%of the unit cell along the field, depending on the precise magnetic
%structure.

\section{Smectic state\label{sec:Smectic-state}}

The anisotropic Wigner crystal variational wavefunction described
above \emph{assumes} that translational symmetry is broken.
However, long-range quantum fluctuations can restore translational
symmetry, either partially or fully, resulting in either a smectic
state (which breaks translational symmetry in only one direction),
a nematic state which breaks rotational symmetry but is
translationally invariant, or an isotropic liquid. Since the
ground state energy is mostly sensitive to \emph{short}-range
correlations, the crystal and the various liquid states can have
close energies. Below, we demonstrate that one can write an
explicit wavefunction describing a smectic state with the same
parametric dependence of the ground state energy on density as
that of the anisotropic WC. In the next section, a ferromagnetic
nematic variational wavefunction will be described.

Let us consider, for simplicity, the case of short-range (contact)
interactions. To construct a wavefunction for the smectic, we consider
a trial Hamiltonian in which the electrons are confined to move along an array of strips %``tubes''
 of width $L_{y}$ with infinitely hard walls separating different strips.
The problem is then reduced to finding the electronic ground state of a strip %tube
with a linear density of $nL_{y}$. The single-particle dispersion
in the strip %tube
is derived in Appendix \ref{sec:App} (Eq.~\ref{dispersion}). The
dispersion of the each transverse sub-band has two degenerate
``valleys'' at $\pm\left(k_{0}+\delta k_{x}^{\star}\right)$, where
$\delta k_{x}^{\star}\sim1/\left(k_{0}L_{y}^{2}\right)$. We will
assume that only the lowest sub-band of the strip is occupied.
(For a fixed $L_y$, this is valid for a sufficiently small
density\cite{comment-lowest-subband}.) Moreover, we may ignore the
``valley'' degeneracy since every electron can be assumed to be in
one of the two valleys, and exchange processes between the valleys
are suppressed. We thus obtain an effective one-dimensional
Hamiltonian for the motion along a strip:

\begin{equation}
H=\sum_{i}\left(\Delta-\frac{\partial_{x}^{2}}{2m^{\star}}\right)+\frac{1}{2}\sum_{i,j}V\left(x_{i}-x_{j}\right),
\end{equation}
where, from Eq.(\ref{dispersion}), $\Delta=A_{1}^{2}/8mk_{0}L_{y}^{4}$
and $m^{\star}\propto m$ ($A_{1}$ is a dimensionless constant).
In the $n\rightarrow0$ limit, the interaction becomes strong compared
with the Fermi energy, and the electrons behave as effectively hard
core particles (independently of their valley index). The system can
be mapped onto a non-interacting spinless fermion problem. The ground
state energy per particle is thus
\begin{equation}
\varepsilon\left(L_{y}\right)=\frac{A_{1}}{8mk_{0}^{2}L_{y}^{4}}+\frac{\left(\pi nL_{y}\right)^{2}}{6m^{*}}.
\end{equation}
 Minimizing this expression with respect to $L_{y}$, we obtain the
ground state energy per particle of the smectic state:

\begin{equation}
\varepsilon_{\mathrm{SM}}\sim\frac{k_{0}^{2}}{m}\left(\frac{n}{k_{0}}\right)^{\frac{4}{3}}.\label{eq:E_SM}
\end{equation}
 The scaling of the energy of the smectic state with $n$ is thus
the same as that of the anisotropic Wigner crystal,
Eq.(\ref{Ekin})\cite{comment-lowest-subband}. The numerical
prefactor, which cannot be determined reliably from such simple
considerations, is therefore important in determining which of
these two states is favored in the $n\rightarrow0$ limit.

In the case of extended interactions which decay with an exponent
$\alpha$, one can use the same variational wavefunction for the smectic,
in which the expectation value of the potential energy is finite,
and then minimize the total energy over $L_{y}$. The calculation
proceeds in essentially the same way as in Sec. \ref{sub:Extended-short-range-interaction},
and we will not repeat the details here. The result is that the parametric
dependence of the smectic variational energy on $n$ is the same as
that of the Wigner crystal, Eq.(\ref{etot}), for \emph{any} $\alpha$.

\section{Ferromagnetic nematic state\label{sec:Ferromagnetic-nematic-state}}

Finally, we consider a complete melting of the anisotropic Wigner
crystal phase, preserving its preferred orientation. This results in
a nematic state. Similarly to the situation in the Wigner crystal
and the smectic states, we expect that in the low-density limit, only
states in the vicinity of two opposite points $\pm\vec{k}_{0}$ on
the ring of minimal dispersion will be occupied. For simplicity, we
will assume the occupation is limited to the vicinity of only \emph{one}
point on the ring, $\vec{k}_{0}$, which makes the nematic state also
ferromagnetic (with an in-plane magnetization perpendicular to $\vec{k}_{0}$).
Such a state is particularly easy to describe within a Hartree-Fock
approximation. We emphasize, however, that a paramagnetic nematic
state is also possible, although it is not easily captured by a simple
variational wavefunction.

The Hartree-Fock analysis of the ferromagnetic nematic state is straightforward,
and is described in Appendix \ref{App:HF}. At a sufficiently low
density, a spontaneous in-plane magnetization develops, and the Fermi
surface becomes asymmetric. At asymptotically low densities, the Fermi
surface becomes an ellipse centered around one of the points on the
minimal dispersion ring in momentum space. The total variational ground
state energy per particle in this limit scales with density as (see
Eq. \ref{eq:E_FM})

\begin{equation}
\varepsilon_{\mathrm{FM}}\sim\begin{cases}
n^{2\left(1-\frac{1}{\alpha}\right)}, & \alpha\le4\\
n^{\frac{3}{2}}, & \alpha>4.
\end{cases}
\end{equation}
Comparing this to Eqs.(\ref{Ekin}),(\ref{etot}), and (\ref{eq:E_SM}), we see
that the ground state energy of the ferromagnetic state is parametrically
smaller than that of either the anisotropic WC or the smectic states
for $\alpha>3$, making it the best candidate for the ground state.
For $2<\alpha\le3$, the scaling of the ground state energy with density
of all three states has the same exponent. The energies differ only
by the prefactor, which cannot be estimated reliably within the simple
variational approach used here.

The ferromagnetic nematic state spontaneously breaks time reversal
($\mathcal{T}$) and rotational symmetry ($\mathcal{R_{\theta}}$)
about the z axis, as well as the mirror symmetry,
$\mathcal{M}_{y}$, for the plane parallel to the ferromagnetic
moment. However, it preserves the product, $\mathcal{TR}_{\pi}$,
of time-reversal and rotation by $\pi$ (hence the name,
``nematic'') and reflection through the plane perpendicular to the
moment, $\mathcal{M}_{x}$. The latter symmetry insures that there
is no out-of-plane component of the magnetization, and no
anomalous Hall effect. Note that, even though this state carries a
finite crystal momentum, it does not carry a finite current
density, as required by a theorem by F.
Bloch\cite{Bloch1933,Bohm1949,comment-current}. There is, however,
a large anisotropy in the in-plane Drude weight (effective mass).
From our Hartree-Fock state (see Appendix \ref{App:HF}), we find
that the anisotropy
%be of the order of $k_0^2/mh^\star$ where $h^\star$ is given by
%Eq.(\ref{eq:hstar}). In the low-density limit, the anisotropy
scales as $n^{2-\frac{4}{\alpha}}$ for $\alpha<4$, and as $n^{-1}$
for $\alpha\ge 4$, in the low-density limit.

The physics behind the formation of the in-plane ferromagnetic
state is similar to the usual Stoner picture for ferromagnetism:
the system gains exchange energy by polarizing, at the expense of
kinetic energy. In a system with Rashba SOC at low density, the
exchange energy gain exceeds the kinetic energy cost due to the
high density of states. In the low-density limit, the Fermi
surface becomes parametrically anisotropic. Qualitatively, the
short-range correlations in this state are similar to those of the
anisotropic Wigner crystal. This explains why these states are
close in energy, at least for $\alpha\le3$. The long-range
correlations, however, are very different: the ferromagnetic state
is a fluid, whereas the Wigner crystal is an insulator.

\section{Phase diagram}

\label{sec:Exp}

%We now discuss the conditions for the formation of the anisotropic
%Wigner crystal.
So far, we have argued that for sufficiently low density and for short-ranged
interactions, the system breaks rotational invariance, going into
either an anisotropic Wigner crystal, smectic, or a nematic state.
We now discuss the global features of the phase diagram as a function
of density and the interaction range. For concreteness, let us discuss
a 2DEG with Rashba SOC with screened Coulomb interactions, where the
screening is from a nearby metallic gate at a distance $\xi$ away.
The effective electron-electron interaction is $V\left(r\right)\approx e^{2}/\kappa r$
for $r\ll\xi$, where $\kappa$ is the dielectric constant of the
surrounding material, and $V\left(r\right)\sim e^{2}\xi^{2}/\kappa r^{3}$
for $r\gg\xi$.

The broken symmetry state forms at a density $n^{\star}$ at which
various scales become comparable to each other. We define a
density $n_1^\star$ at which the energy of the broken symmetry
state per particle is comparable to that of the uniform Fermi
liquid state (which is dominated by Coulomb energy):
\begin{equation}
\frac{e^{2}\xi
n_1^{\star}}{\kappa}\sim\frac{1}{m}\left(m\frac{e^{2}\xi^{2}}{\kappa}\right)^{\frac{2}{3}}\left(n_1^{\star}\right)^{\frac{4}{3}},\label{eq:EqEn}
\end{equation}
 where $e$ is the electron charge, and $\kappa$ is the dielectric constant
of the host material. On the left hand side we have used that, for
short range interactions, the potential energy of the uniform
state scales linearly with density (assuming that
$(n_1^\star)^{-\frac{1}{2}}\gg \xi$), and on the right hand side
we have used Eq.(\ref{etot}) for the energy of the broken symmetry
state with $V_{0}\sim e^{2}\xi^{2}/\kappa$. Note that the energies
of all the different candidate states are \emph{the same} up to a
numerical prefactor in the case $\alpha=3$; compare
Eqs.(\ref{etot}) and (\ref{eq:E_FM}). This gives $n_1^{\star} \sim
\left(a_{0}\xi\right)$, where $a_{0}\equiv\kappa/me^{2}$ is the
effective Bohr radius.

In addition, %for consistency, the
we define a density $n_2^{\star}$ %must be low enough that
at which the inter-electron distance is comparable to the
screening length $\xi$, and a density $n_3^\star$ at which the
Fermi wavevector is comparable to $k_0$.
%\mpar{ok?}distance is larger than $\xi$, and %the inter-electron spacing
%is of the order of $1/k_{0}$ or larger.
These characteristic densities are given by
$n^{\star}<n_{2}^{\star}\equiv1/\xi^{2}$ and
$n^{\star}<n_{3}^{\star}\equiv k_{0}^{2}$. The strongly
anisotropic states are favored
%The anisotropic Wigner crystal occurs therefore
at densities which satisfy
$n<n^{\star}=\min[n_{1}^{\star},n_{2}^{\star},n_{3}^{\star}]$.

%%%%%%%%%%%%%%%%%%%%%%%%%%%%%%%%%%%%%%%%%%%%%%%%%%%%%%%%%%%%%%%%
\begin{figure}
\includegraphics[scale=0.35]{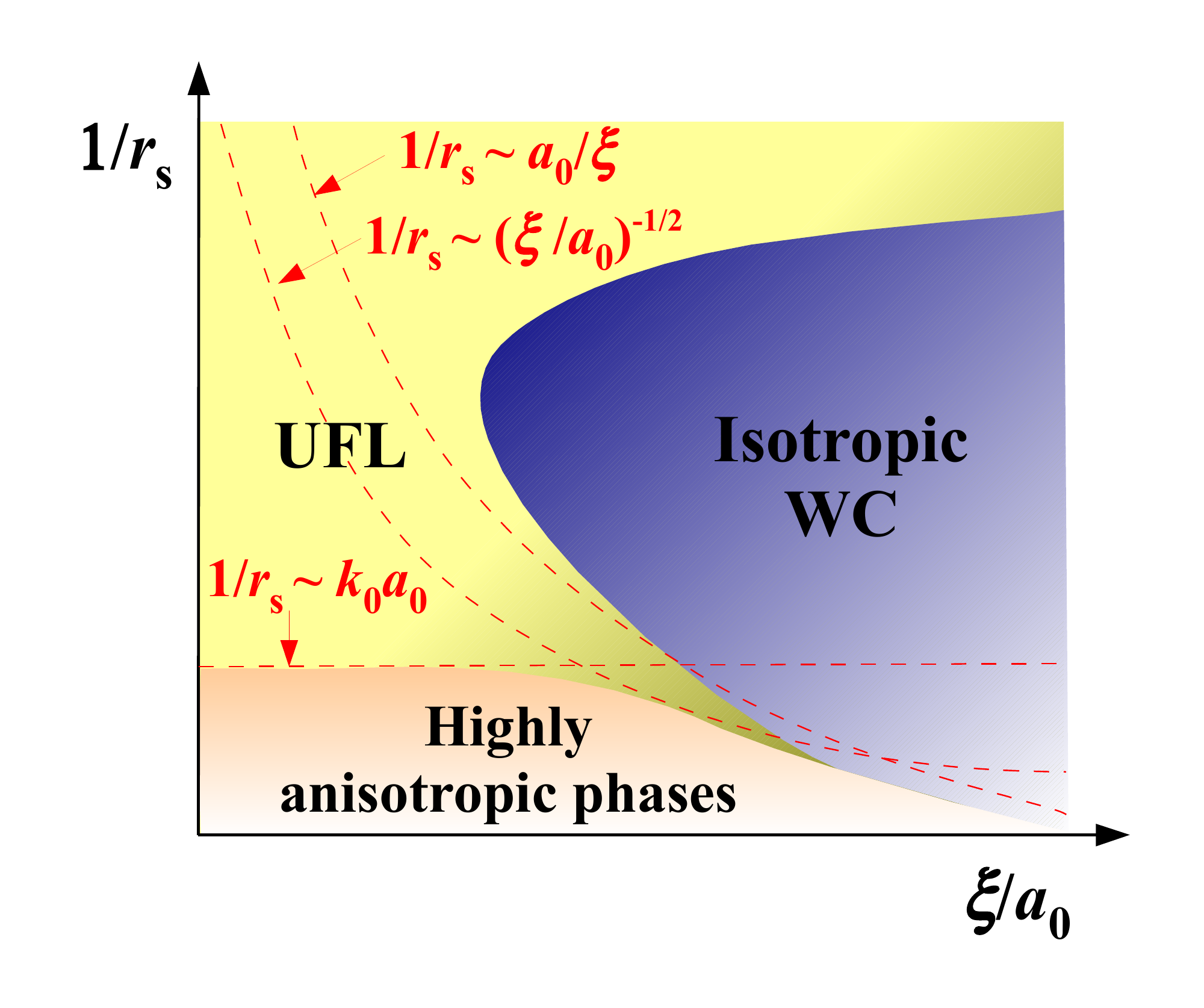}

\caption{Sketch of the phase diagram of a 2DEG with Rashba SOC, as
a function of $1/r_{s}\equiv a_{0}\left(\pi n\right)^{1/2}$ and
the screening length $\xi$, where $a_{0}=\kappa/me^{2}$
is the Bohr radius. The three regions correspond to %phases appearing here are:
the uniform Fermi Liquid (UFL), the isotropic Wigner Crystal (WC),
and a phase featuring a high degree of anisotropy, which can be
either an anisotropic Wigner crystal, a smectic, or a nematic. The
red dashed lines correspond to $r_{s}\sim\xi/a_{0}$, where the
crossover between effectively short-range (screened) and
long-range Coulomb interactions occurs; $1/r_{s}\sim k_{0}a_{0}$,
where the SOC length scale becomes comparable to the
inter-particle distance; and $1/r_{s}\sim(\xi/a_{0})^{-1/2}$,
where the energies of the uniform Fermi liquid and the anisotropic
phases become comparable, see Eq.(\ref{eq:EqEn}).}

\label{fig:PD}
\end{figure}

%%%%%%%%%%%%%%%%%%%%%%%%%%%%%%%%%%%%%%%%%%%%%%%%%%%%%%%%%%%%%%%%

We can now speculate about the structure of the zero temperature
phase diagram of a 2DEG with Rashba SOC, sketched in
Fig.~\ref{fig:PD}, as a function of the dimensionless
inter-electron spacing $r_{s}\equiv\left(\pi
na_{0}^{2}\right)^{-1/2}$ and the screening length $\xi$. Let us
consider large $\xi$, for which the Coulomb interactions are
effectively unscreened. Then, a Wigner crystal with hexagonal
symmetry forms when $r_{s}>r_{s,c}\approx35$.\cite{Tanatar1989}
Imagine that we start deep in the Wigner crystal phase, with
arbitrarily large $r_{s}$ and $\xi$. Upon decreasing $\xi$, while
keeping $r_{s}$ fixed, eventually we reach $\xi\lesssim
r_{s}a_{0}$, where the interactions are effectively short-ranged
and the kinetic energy becomes important. At some point along this
path, we expect a phase transition from the hexagonal Wigner
crystal to one of the phases of lower rotational symmetry: either
an anisotropic Wigner crystal, a smectic, or a nematic, which can
also be ferromagnetic. Which of these phases is realized cannot be
determined reliably on the basis of the present analysis.

At higher densities, such that $r_{s,c}<r_{s}<1/(k_{0}a_{0})$, the
SOC can essentially be ignored. Then, upon decreasing $\xi$ from
the Wigner crystal, we expect a %(possibly first order)
transition to a Fermi liquid. The reentrant tip of the Wigner
crystal phase originates from the same physical reasoning as that
described in Ref~\onlinecite{Spivak2004}.

As drawn, a sliver of UFL is shown between the isotropic WC and
broken rotational symmetry phases. Such a region may or may not
exist, depending on details of the numerical factors which are
beyond the scope of the calculation here.

%For $r_{s}\ll\xi/a_{0}$, the Coulomb interaction is effectively
%unscreened, and the system forms an isotropic Wigner crystal when
%$r_{s}>r_{s,c}\approx35\kappa$.\cite{Tanatar1989} An anisotropic
%Wigner crystal forms below a critical density
%%SAK in which
%Eq.~(\ref{eq:EqEn}) is satisfied, given that the inter-particle
%distance is sufficiently larger than the SOC length $1/k_{0}$
%(otherwise, SOC can be ignored), and that $r_{s}\gtrsim\xi/a_{0}$
%(such that the interactions become effectively short-ranged). Of
%course, a more rigorous analysis would be required to confirm this
%phase diagram. Note also that in drawing Fig. \ref{fig:PD}, it has
%been assumed that $1/r_{s,c}>k_{0}a_{0}$ (which may or may not
%hold in a particular realization).

\section{Possible realizations}

\label{sec:real}

The physics described here could be relevant to electrons in two-dimensional
heterostructures which lacks inversion symmetry, such as GaAlAs quantum
wells. The magnitude of the Rashba spin-orbit coupling in these systems,
however, is rather small. As discussed above, a necessary condition
for realizing the phases with broken rotational symmetry %anisotropic Wigner crystal phase
is $r_{s}\gtrsim1/k_{0}a_{0}$, or equivalently, $k_{F}\lesssim
k_{0}$; in typical GaAlAs quantum wells, $k_{0}/k_{F}\sim0.5$ or
less, and the characteristic energy scale of the SOC, $E_{0}$ is
at most of the order of a few degrees
Kelvin\cite{WinklerBook,Nitta1997}. More promising systems are
surface states of heavy metal surface alloys. For instance, the
boundary between Bi$_{x}$Pb$_{1-x}$ and Ag(111) supports a surface
state with very strong Rashba SOC, with
$k_{0}\approx2\mathrm{nm}^{-1}$ and
$E_{0}\approx0.1\mathrm{eV}$.\cite{Ast2007,Meier2009} Moreover, it
was demonstrated that by varying $x$, the Fermi level of the
surface state can be tuned to be lower than $E_{0}$.\cite{Ast2008}
The Coulomb interactions on the surface are naturally screened by
the metallic bulk\cite{comment-metallic-surface}. Detecting the
broken rotational symmetry on the surface poses a challenge,
because transport measurements would be dominated by the bulk. One
possibility is to look for signatures of anisotropy in the
finite-frequency response, e.g. in the optical conductivity,
assuming that the anisotropic domains can be aligned (e.g., by
application of an in-plane magnetic field). Scanning tunneling
microscopy can be done on metallic surface alloys\cite{Ast2007a},
and used to detect anisotropy in the electronic structure.
Finally, magnetic spectroscopy can be used to detect the
ferromagnetic state, which has a large in-plane moment. Such
measurements have recently been done\cite{Bert2011} on the
conducting interface between LaAlO$_{3}$ and SrTiO$_{3}$, and
indeed, large in-plane moments were found. Whether these are
related to the mechanism described in this paper remains to be
seen.

It has been proposed\cite{Jung2011} that similar physics can arise
lightly doped bilayer graphene with a perpendicular electric
field, in which the single particle dispersion has a minimum on a
ring in k-space, even without SOC. %This system was recently
%predicted to host a broken symmetry \mpar{mention current?}state
%which is similar in nature to the ones found here\cite{Jung2011},
%although it has been proposed that this state also carries a
%finite persistent current.
In bilayer graphene the single particle
dispersion is valley and spin degenerate. The ground state is
likely to have additional broken symmetries, lifting these
degeneracies.

It is interesting to note that the considerations which lead to
broken rotational symmetry at low densities are independent of the
particle statistics, and are thus valid for two-component bosons
with effective (isotropic) Rashba-like spin orbit interactions.
Recently, various techniques were proposed to realize effective
SOC in systems of trapped ultracold
atoms\cite{Stanescu2007,Campbell2011}. The properties of such
systems have been the subject of intense
study\cite{Wu2011,Stanescu2008,Wang2010,Gopalakrishnan2011}.
%It seems that
%The anisotropic Wigner crystal phase
Highly anisotropic phases may be accessible in such systems
%a good candidate
at sufficiently low densities. Indeed, it was found\cite{Gopalakrishnan2011}
that the ground state breaks rotational symmetry at low densities,
and that the ground state energy per particle scales as $n^{4/3}$
in the limit $n\rightarrow0$, consistently with our results for the
anisotropic WC and smectic phases with contact interactions.

\section{Conclusions}

\label{sec:conclude}

In the presence of strong Rashba SOC, even short-range electron-electron
interactions become important. As a result, the system is expected
to form a broken symmetry state at low enough densities. In this work,
we have shown that for sufficiently short-range interactions, states
which break rotational symmetry are favored in the low-density limit.
This is a consequence of the fact that the single-particle dispersion
has a minimum on a ring of finite radius in $k$-space, rather than
at a single point. This physics is not limited to the case of Rashba
SOC; for instance, a similar situation arises in spin-imbalanced fermionic
superfluids with no SOC, in which the majority-spin quasiparticles
have a dispersion which is minimal near $k_{F}$,\cite{Yang2006}
or in bilayer graphene with a transverse electric field\cite{Jung2011}.

%The approach taken here is too simple to answer some important, more detailed
%questions about the anisotropic Wigner crystal, such as what is its
%precise crystal structure and possible magnetic order. Nevertheless,
We believe that the variational wavefunctions and physical
arguments proposed above capture the correct \emph{scaling} of the
ground state energy, which is found to be parametrically lower
than other states (e.g.~a uniform Fermi liquid or an isotropic
Wigner crystal). However, this approach is too crude to answer
some important, more detailed questions, such as discriminating
between the different broken symmetry states considered here. For
sufficiently short-range interactions (which fall off with an
exponent larger than 3) a nematic ferromagnetic state has a
parametrically lower energy than all the other states considered
here, and is therefore the best candidate for the ground state. It
is not clear, at this point, whether a \emph{non-magnetic} nematic
state is a competitor or not; such a state is harder to capture
within a simple variational approach. More detailed calculations
will be needed to determine the phase diagram for interactions
which fall off with distance with a power of 3 or less. For
instance, it may be interesting to treat this problem in an
unrestricted Hartree-Fock approximation, which can be used to
systematically improve the variational wavefunctions used in this
work.

Edge states of surface alloys, such as the one discovered by Ast et
al.\cite{Ast2007}, seem to be promising candidates to realize the
anisotropic Wigner crystal phase, since they combine extremely strong
Rashba SOC, a tunable Fermi energy, and screening due to the nearby
metal. In a real system, however, disorder will inevitably play a
major role. As a result, both the positional and orientational order
are expected to be short-ranged. To detect the broken symmetry state
on the surface, one can either resort to local probes (such as scanning
tunneling spectroscopy), or find a way to align the orientational
domains, e.g. by an in-plane magnetic field.

\acknowledgements{We thank E. I. Rashba, E. Demler, A. Amir, I.
Neder, and B. I. Halperin for useful discussions. We are
particularly indebted to Srinivas Raghu for his part in
stimulating this project. This work was supported by the NSF under
grants DMR-0757145, DMR-0705472 (E.~B.), DMR-090647 and
PHY-0646094 (M.~S.~R.), and by DOE grant \# AC02-76SF00515 at
Stanford (S.~A.~K.). M.R. thanks the IQOQI for their hospitality.
Work at the IQOQI was supported by SFB FOQUS through the Austrian
Science Fund, and the Institute for Quantum Information.}

\bigskip{}

\appendix
%dummy comment inserted by tex2lyx to ensure that this paragraph is not empty
%dummy comment inserted by tex2lyx to ensure that this paragraph is not empty
%dummy comment inserted by tex2lyx to ensure that this paragraph is not empty
%dummy comment inserted by tex2lyx to ensure that this paragraph is not empty
%dummy comment inserted by tex2lyx to ensure that this paragraph is not empty
%dummy comment inserted by tex2lyx to ensure that this paragraph is not empty

\section{Rashba particle in a box}

\label{sec:App}

The arguments presented in this paper rely on the solution of the
quantum mechanical problem of a single particle with Rashba SOC in
a rectangular box of size $L_{x}\times L_{y}$ with infinite potential
walls. While the corresponding problem without spin-orbit coupling
is trivial, with spin-orbit coupling the problem of boundary-condition
matching with the multi-component wavefunction is highly non-trivial.
The reason the problem with SOC is more difficult is that in this
case, the Hamiltonian is no longer separable (i.e., it cannot be written
as a sum of two commuting terms, one of which depends only on the
$x$ coordinate and the other on $y$). A \emph{circular} well can
be solved exactly\cite{Tsitsishvili2004}, thanks to its rotational
invariance. %EB and we do not know how to solve it exactly.

In this Appendix, we combine several approaches to deduce the asymptotic
form of the ground state energy in the anisotropic, low--density limit,
Eq.~(\ref{Enl}) of the main text. We first solve the problem exactly
in the $L_{x}\rightarrow\infty$ limit (keeping $L_{y}$ fixed), and
then provide an argument which yields the form of the leading corrections
for finite $L_{x}$. Finally, we present numerical results supporting
the analytical arguments.

\subsection{Solution in the $L_{x}\rightarrow\infty$ limit}

In the limit $L_{x}\rightarrow\infty$, the problem becomes translationally
invariant in the $x$ direction. The eigenfunctions then take the
form
\begin{equation}
\psi(x,y)=e^{ik_{x}x}\varphi(y),
\end{equation}
 where $\psi$, $\varphi$ are two--component spinors. We choose coordinates
such that the walls are at $y=\pm L_{y}/2$. Then $\varphi(y)$ satisfies
the boundary conditions
\begin{equation}
\varphi(\pm L_{y}/2)=\left(\begin{array}{c}
0\\
0
\end{array}\right)\text{.}\label{BC}
\end{equation}
 Seeking a solution with energy $E=\varepsilon$, we get
that the wavevector modulus $k=\sqrt{k_{x}^{2}+k_{y}^{2}}$
satisfies
\begin{equation}
\varepsilon=\frac{\left(k-k_{0}\right)^{2}}{2m}\text{.}\label{eps}
\end{equation}
 Fixing $k_{x}$, we %get that there are
find four allowed values of $k_{y}$,
which we denote by $\pm k_{y,\pm}$: % where:
\begin{equation}
k_{y,\pm}=\sqrt{\left(k_{0}\pm\sqrt{2m\varepsilon}\right)^{2}-k_{x}^{2}}\text{.}\label{ky}
\end{equation}
 Note that $k_{y,\pm}$ can be imaginary.

\begin{figure}[b]
 \includegraphics[width=0.45\textwidth]{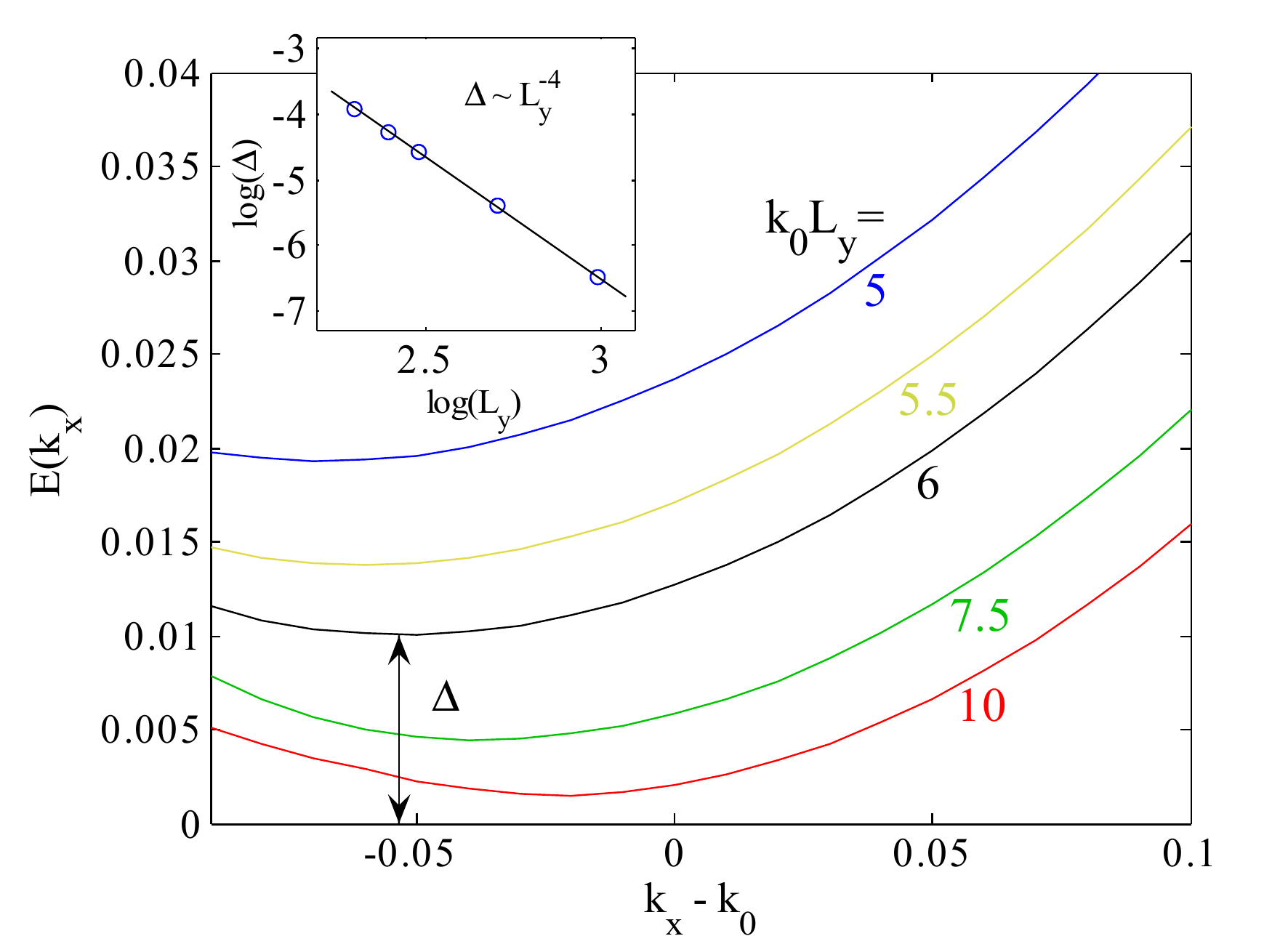}
\caption{Dispersion of the lowest subband for infinite strips of
varying width $L_y$. The inset shows the dispersion minimum
$\Delta$ \textbf{\textit{vs.}} $L_{y}$ on a log-log plot, showing
good agreement with $\Delta\sim1/L_{y}^{4}$.}

\label{fig:dispersion}
\end{figure}

The wavefunction for the transverse motion takes the form
\begin{equation}
\varphi(y)=\sum_{\eta_{1},\eta_{2}=\pm}a_{\eta_{1}\eta_{2}}e^{i\eta_{1}k_{y,\eta_{2}}y}\left(\begin{array}{c}
e^{-i\eta_{1}\theta_{\eta_{2}}/2}\\
ie^{i\eta_{1}\theta_{\eta_{2}}/2}
\end{array}\right)\text{,}\label{phi}
% \varphi(y)=\sum_{\varepsilon_{1},\varepsilon_{2}=\pm}a_{\varepsilon_{1}\varepsilon_{2}}e^{i\varepsilon_{1}k_{y}^{\varepsilon_{2}}y}\left(\begin{array}{c}
% e^{-i\varepsilon_{1}\theta_{\varepsilon_{2}}/2}\\
% ie^{i\varepsilon_{1}\theta_{\varepsilon_{2}}/2}
% \end{array}\right)\text{,}\label{phi}
\end{equation}
 where
\begin{equation}
e^{i\theta_{\eta_{1,2}}}=\frac{k_{x}+ik_{y,\eta_{1,2}}}{k}\label{theta}
\end{equation}
 and $a_{\eta_{1},\eta_{2}}$ ($\eta_{1},\eta_{2}=\pm$)
are coefficients which are determined by the boundary conditions.
We may reduce the number of coefficients by using symmetry. Under
reflection, $y\rightarrow-y$ the spinor $\varphi(y)$ transforms
as $\varphi(y)\rightarrow\sigma_{y}\varphi(-y)$. Requiring that the
wavefunctions are either even or odd under reflection gives
\begin{equation}
a_{\eta_{1},\eta_{2}}=\pm a_{-\eta_{1},\eta_{2}}
%a_{\varepsilon_{1},\varepsilon_{2}}=\pm a_{-\varepsilon_{1},\varepsilon_{2}}
\end{equation}
 Imposing the boundary condition, Eq.~(\ref{BC}), on the wavefunction
in Eq.~(\ref{phi}), %then
for the even sector gives the following (implicit) equation for
$\varepsilon$:
\begin{widetext}
\begin{equation}
\cos\left(\frac{k_{y,+}L_{y}}{2}-\frac{\theta_{+}}{2}\right)\cos\left(\frac{k_{y,-}L_{y}}{2}+\frac{\theta_{-}}{2}\right)-\cos\left(\frac{k_{y,+}L_{y}}{2}+\frac{\theta_{+}}{2}\right)\cos\left(\frac{k_{y,-}L_{y}}{2}-\frac{\theta_{-}}{2}\right)=0\text{,}\label{sec}
\end{equation}
 \end{widetext}where $k_{y,\pm}$, $\theta_{\pm}$ are given by
Eqs.~(\ref{ky}) and (\ref{theta}). For the odd sector, we get an
identical equation with $\cos$ replaced by $\sin$. Equation (\ref{sec})
determines the dispersion $\varepsilon(k_{x})$. The dispersion of
the lowest subband, near $k_{x}=k_{0}$, %the  of the lowest subband are
is shown in Fig.~\ref{fig:dispersion}, for widths $k_{0}L_{y}/2$
ranging from 10 to 20.
%%%%%%%%%%%%%%%%%%%%%%%%%%%%%%%%%%%%%%%%%%%%%%%%%%%%%%%%%%%%%%%%%%%%%%
%%%%%%%%%%%%%%%%%%%%%%%%%%%%%%%%%%%%%%%%%%%%%%%%%%%%%%%%%%%%%%%%%%%%%%

The form of the low energy dispersion can be deduced from
Eq.~(\ref{sec}) as follows. To lowest order in $\varepsilon$ and
$\delta k_{x}\equiv k_{x}-k_{0}$, this equation depends on
$\varepsilon$, $\delta k_{x}$, and $L_{y}$ through the factors
$k_{y,\pm} L_{y}:$
\begin{eqnarray}
k_{y,\pm}L_y & = & \sqrt{\left(k_{0}\pm\sqrt{2m\varepsilon}\right)^{2}-\left(k_{0}+\delta k_{x}\right)^{2}}L_{y}\nonumber \\
 & \approx & \sqrt{\pm2k_{0}\sqrt{2m\varepsilon}-2k_{0}\delta k_{x}}L_{y}.
\end{eqnarray}
 Therefore Eq.~(\ref{sec}) has the functional form
\begin{equation}
F\left(2k_{0}\sqrt{2m\varepsilon}L_{y}^{2},2k_{0}\delta k_{x}L_{y}^{2}\right)=0,
\end{equation}
 where $F$ is some function of two variables. Close to the band minimum,
the dispersion has the following form:
\begin{equation}
2k_{0}\sqrt{2m\varepsilon}L_{y}^{2}=A_{1}+A_{2}\left[2k_{0}\delta k_{x}L_{y}^{2}-A_{3}\right]^{2},
\end{equation}
 where $A_{1,2,3}$ are dimensionless constants. Therefore
\begin{eqnarray}
\varepsilon\left(\delta k_{x}\right) & \approx & \frac{1}{8mk_{0}^{2}}\left[\frac{A_{1}}{L_{y}^{2}}+A_{2}\left(2k_{0}\delta k_{x}-\frac{A_{3}}{L_{y}^{2}}\right)^{2}L_{y}^{2}\right]^{2}\nonumber \\
 & \approx & \frac{1}{8mk_{0}^{2}}\frac{A_{1}^{2}}{L_{y}^{4}}+\frac{A_{1}A_{2}}{m}\left(\delta k_{x}-\delta k_{x}^{\star}\right)^{2},\label{dispersion}
\end{eqnarray}
 where $\delta k_{x}^{\star}=A_{3}/2k_{0}L_{y}^{2}$. We see the dispersion
minimum $\Delta$ scales as $1/L_{y}^{4}$, while the effective mass
in the $x$ direction is $L$ independent. We confirm the relation
$\Delta\sim1/L_{y}^{4}$ by solving Eq.~(\ref{sec}) numerically,
see inset of Fig.~\ref{fig:dispersion}.

\subsection{Extension to Finite $L_{x}$}

\begin{figure}[t]
 \includegraphics[width=0.48\textwidth]{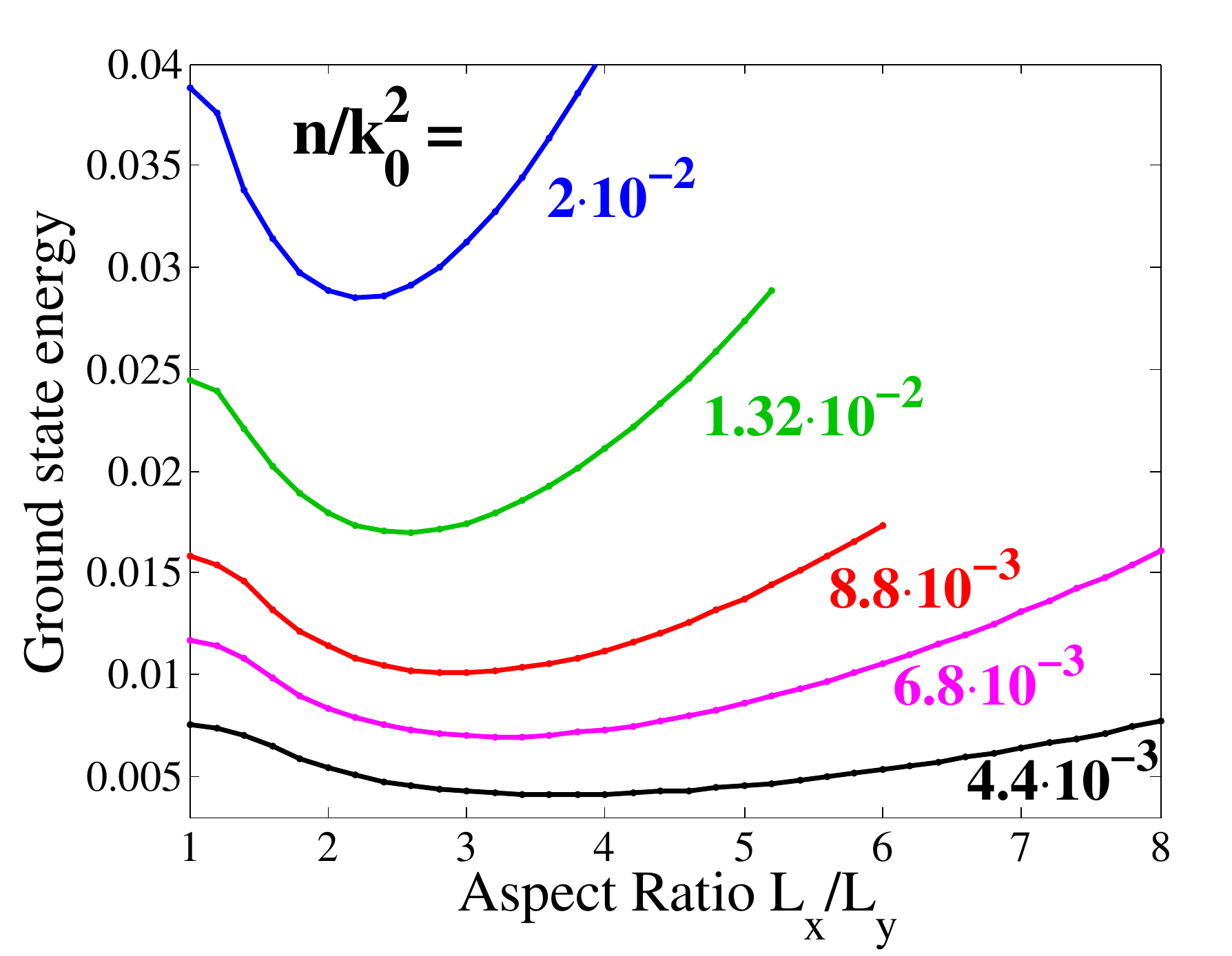}
\caption{Numerical results for the ground state energy vs. aspect ratio $L_{x}/L_{y}$
of a particle with Rashba spin orbit coupling in a rectangular infinite
potential well of size $L_{x}\times L_{y}$, for several values of
the density $n$ (measured in units of $1/k_{0}^{2}$). We have set
$2m=1$.}

\label{fig:E_l}
\end{figure}

\begin{figure}[t]
 \includegraphics[width=0.5\textwidth]{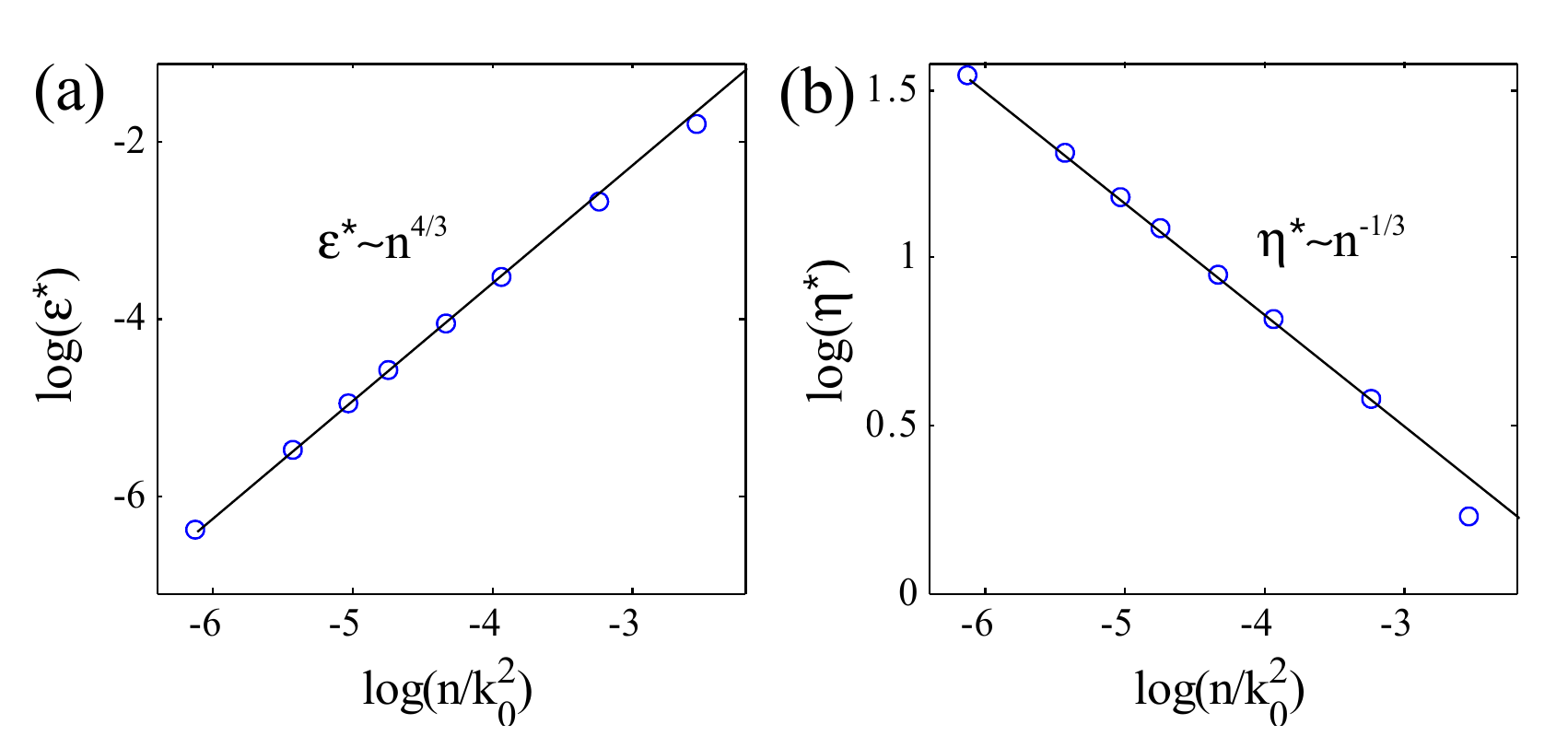}
\caption{(a) Minimal ground state energy, $\varepsilon^{\star}$, and (b) optimal
aspect ratio, $\eta^{\star}$, as a function of the density, on a
log-log scale. The data is consistent with the analytical predictions:
$\varepsilon^{\star}\sim n^{4/3}$ and $\eta^{\star}\sim n^{-1/3}$
at low densities.}

\label{fig:loglog}
\end{figure}

Next, we consider that $L_{x}$ is finite, still assuming that $L_{x}\gg L_{y}$.
In this case, the solution is complicated by multiple reflections
from the boundaries. Nevertheless, we can deduce the form of the ground
state energy as a function of $L_{x}$ and $L_{y}$ as follows. For
large $L_{x}$, we assume that the solution is largely composed of
traveling wave states near the bottom of the dispersion given by Eq.~(\ref{dispersion}),
which are reflected back and forth from the two edges. Consider a
right-moving wave with momentum $k_{x}=k_{0}+\delta k_{x}$. This
state can \emph{only} be reflected to the state $k_{x}^{\prime}=k_{0}-\delta k_{x}+2\delta k_{x}^{\star}$,
where $\delta k_{x}^{\star}$ is defined below Eq.~(\ref{dispersion}).
Note that time reversal symmetry prohibits scattering to the other
left-moving solution with momentum $-k_{0}-\delta k_{x}$, since this
state is the Kramer's partner of the original incoming wave. % and time reversal symmetry prohibits scattering between Kramer's pairs.
The eigenstates of the system are determined by the requirement that
the phase acquired over one period is a multiple of $2\pi$:
\begin{equation}
2\left(\delta k_{x}-\delta k_{x}^{\star}\right)L_{x}+\phi=2\pi j\text{,}
\end{equation}
 where $j$ is an integer, and $\phi=\phi_{1}+\phi_{2}$ is the sum
of the two (unknown) phase shifts $\phi_{1}$, $\phi_{2}$ associated
with reflections from the two ends. The total phase shift $\phi$
is a function $\delta k_{x}$, $L_{y}$ and $k_{0}$. However, in
the limit $\delta k_{x}L_{y}\rightarrow0$, $1/k_{0}L_{y}\rightarrow0$,
we assume that we can replace $\phi\left(\delta k_{x}L_{y},1/k_{0}L_{y}\right)$
by a constant $\phi\left(\delta k_{x}L_{y}\rightarrow0,1/k_{0}L_{y}\rightarrow0\right)\equiv\phi_{0}$.
We then find that the ground state is given by $\delta k_{x}-\delta k_{x}^{\star}=(2\pi j_{\min}-\phi_{0})/2L_{x}$,
where $j_{\min}$ is an integer chosen to minimize the energy below.
Inserting this into Eq.~(\ref{dispersion}), we finally get
\begin{equation}
\varepsilon\left(L_{x},L_{y}\right)\approx\frac{1}{8mk_{0}^{2}}\frac{A_{1}^{2}}{L_{y}^{4}}+\frac{A_{1}A_{2}\left(\pi j_{\min}-\phi_{0}/2\right)^{2}}{mL_{x}^{2}}\text{.}\label{eLxLy}
\end{equation}
 Substituting $n=1/(L_{x}L_{y})$ and $\eta=L_{x}/L_{y}$, we obtain
Eq.~(\ref{Ekin}).

\subsection{Numerical solution}

In order to verify the assumptions that lead to Eq.~(\ref{eLxLy}),
we have numerically calculated the ground state wavefunction of a
Rashba particle in a box. This is done using a generalization of
the {}``plane wave decomposition'' technique described in
Refs.~\onlinecite{Heller1984,Li1998}. The solution is written as a
superposition of eigenstates of the free Hamiltonian, which are
plane waves with wavevectors satisfying Eq.~(\ref{eps}) for some
value of $\varepsilon$. The coefficients of the different plane
waves are determined by requiring that both components of the
wavefunction vanish on a set of points evenly distributed along
the boundary (in this case, an $L_{x}\times L_{y}$ rectangle), and
that the first component of the wavefunction spinor is equal to 1
at an arbitrary point in the interior. The sum of the squares of
the wavefunction (the {}``tension'') at a different set of points
on the boundary is then calculated. The eigenvalues $\varepsilon$
are identified as the minima of the tension. The eigenvalues can
be determined with an accuracy of 1\% or better. We have tested
the technique by calculating the eigenenergies of a Rashba
particle in a circular well, and found excellent agreement with
the exact results\cite{Tsitsishvili2004}.

Fig.~\ref{fig:E_l} shows the ground state energy vs. the aspect
ratio $\eta=L_{x}/L_{y}$ for a few values of the density $1/n=L_{x}L_{y}$.
The ground state energy is minimal for $\eta\ne1$. We found this
behavior for $n\lesssim0.08k_{0}^{2}$; for higher densities (smaller
box area), the minimum energy occurs at $\eta=1$. In Fig.~\ref{fig:loglog}
we show the minimal ground state energy $\varepsilon^{\star}$ and
the corresponding aspect ratio $\eta^{\star}$ for densities ranging
from $n/k_{0}^{2}=8\times10^{-2}$ to $4\times10^{-3}$. For low densities,
the optimal ground state energy and aspect ratio follow $\varepsilon^{\star}\sim n^{4/3}$
and $\eta^{\star}\sim n^{-1/3}$, in agreement with Eqs.~(\ref{Asp}),(\ref{Ekin}),
and (\ref{eLxLy}).

\section{Hartree-Fock description of the ferromagnetic nematic\label{App:HF}}

Within the Hartree-Fock approximation, we replace the full Hamiltonian $H$, Eq.(\ref{H}), by the following mean-field Hamiltonian $\mathcal{H}$:

\begin{equation}
\mathcal{H}  =  \sum_{j}\left(\frac{1}{2m}\left[-\nabla_{j}^{2}-\frac{2k_{0}}{i}(\nabla_{j}\times{\hat{z}})\cdot\vec{\sigma}_{j}\right]\label{eq:Hmf}
 - \mu - \frac{1}{2}\vec{h}\cdot\vec{\sigma_{j}}\right),
\end{equation}
 where $\vec{h}$ is a spontaneous Zeeman field, to be determined
self-consistently, and $\mu$ is the chemical potential. We proceed
by minimizing the expectation value of the full Hamiltonian within
the ground state of the mean-field Hamiltonian, Eq.(\ref{eq:Hmf}).

Let us focus on the case of an \emph{in-plane} Zeeman field $\vec{h}$.
(We will later argue that an out-of-plane $\vec{h}$ is not energetically
favorable.) Without loss of generality, we assume that $\vec{h}=h\mathbf{\hat{x}}$.
Then, the lower-branch single particle dispersion obtained by diagonalizing
Eq.(\ref{eq:Hmf}) is
\begin{equation}
\varepsilon_{k}=\frac{k^{2}}{2m}-\mu-\frac{k_{0}}{m}\sqrt{k_{x}^{2}+\left(k_{y}+\frac{mh}{2k_{0}}\right)^{2}}.\label{eq:MF-dispersion}
\end{equation}
 The minimum of the dispersion is obtained for $\vec{k}=k_{0}\mathbf{\hat{y}}$.
In the low-density limit, only states close to the minimum are
occupied. We therefore expand the dispersion around the minimum,
using polar coordinates, $k_{x}=k\sin\theta$ and
$k_{y}=k\cos\theta$ where $k=k_{0}+\delta k$, to leading order in
$\delta k$ and $\theta$. This gives
\begin{equation}
\varepsilon_{k}\approx-\frac{k_{0}^{2}}{2m}-\frac{1}{2}h^{x}-\mu+\frac{\delta k^{2}}{2m}+\frac{1}{4}\frac{k_{0}^{2}h}{(k_{0}^{2}+\frac{m}{2}h)}\theta^{2}.
\end{equation}
 %%%%%%%%%%%%%%%%%%%%%%%%%%%%%%%%%%%%%%%%%%%%%%%%%%%%%%%%%%%%%%%%%%%%%%
\begin{figure}[t]
 \includegraphics[width=0.45\textwidth]{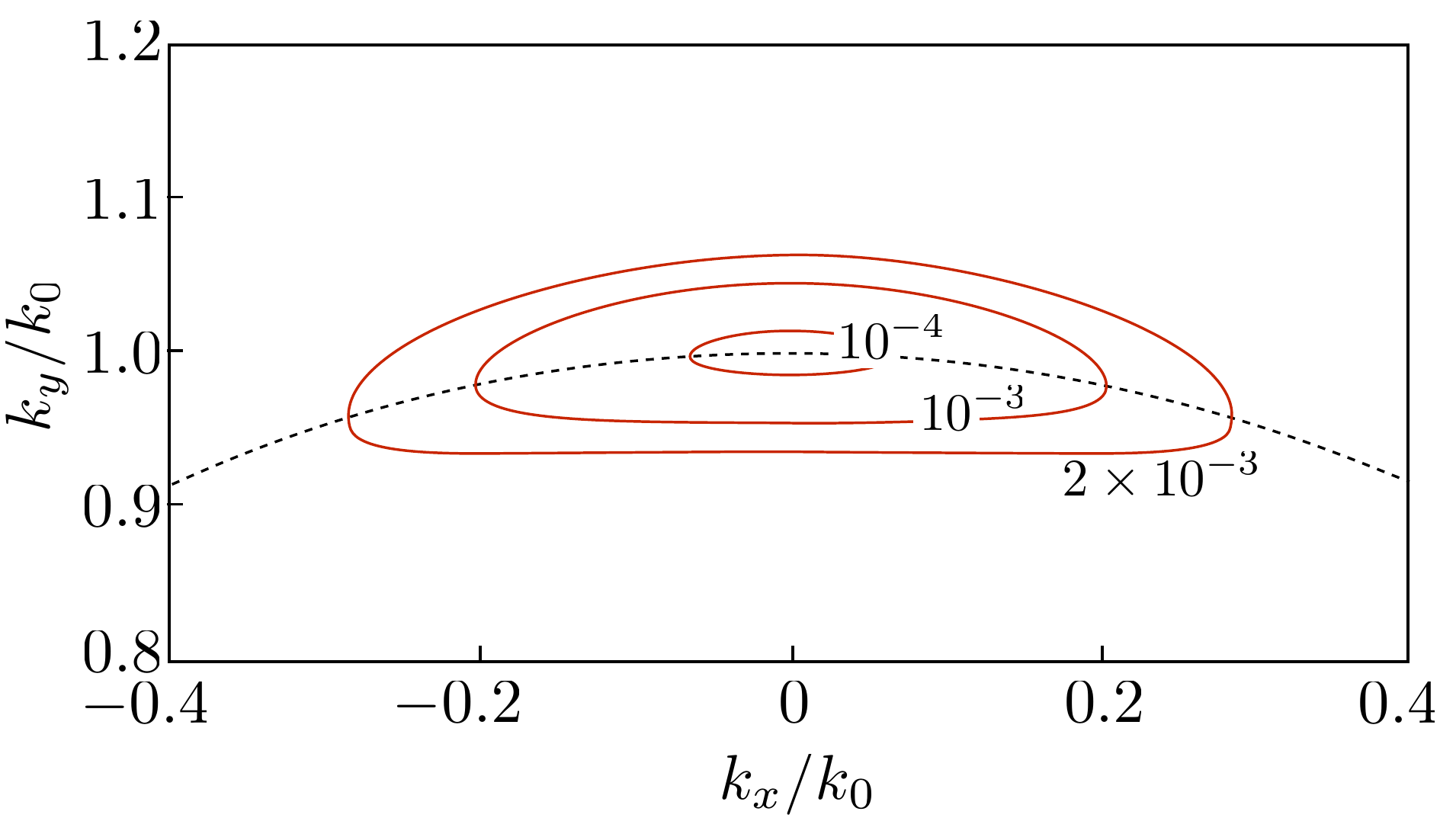}
\caption{ Constant energy contours of the dispersion of the
mean-field Hamiltonian, Eq. \ref{eq:MF-dispersion}, around the
minimum at $\vec{k} = k_0\mathbf{\hat{y}}$. Numerical labels
indicate energy values above the minimum, measured in units of
$k_0^2/(2m)$. The dashed line indicates the circle $k_x^2 + k_y^2
= k_0^2$.} \label{fig:bananas}
\end{figure}
%%%%%%%%%%%%%%%%%%%%%%%%%%%%%%%%%%%%%%%%%%%%%%%%%%%%%%%%%%%%%%%%%%%%%%

The Fermi surface is thus approximately an ellipse centered around
$k_{0}\mathbf{\hat{y}}$ (see Fig.\ref{fig:bananas}). Denoting the
Fermi energy measured relative to the dispersion minimum
$\varepsilon_{0}=\frac{k_{0}^{2}}{2m}+\frac{1}{2}h^{x}+\mu$ as
$\varepsilon_F$, we get that the density and Fermi energy are related by %as a function of density is
\begin{equation}
n=\frac{m}{2\pi}\sqrt{1+\frac{2k_{0}^{2}}{mh}}\varepsilon_{F}.
\end{equation}
 The expectation value of the kinetic energy per particle in the ground
state of $\mathcal{H}$ (measured relative to the band minimum,
$\varepsilon_0$) is

\begin{align}
K & =\langle\mathcal{H}+\frac{1}{2}\vec{h}\cdot\vec{\sigma_{j}}\rangle+\frac{k_{0}^{2}}{2m}+\mu\approx\left\langle \frac{\delta k^{2}}{2m}\right\rangle \nonumber \\
 & \approx\frac{\pi n}{2}\sqrt{\frac{h}{2mk_{0}^{2}}},\label{eq:K}
\end{align}
 where we kept only the leading order term in $\frac{mh}{k_{0}^{2}}$.
Next, we calculate the potential energy. Assuming that
$V\left(r\right)\sim r^{-\alpha}$ at long distances, its Fourier
transform has the following form for small momentum transfer:

\begin{equation}
\tilde{V}\left(q\right)\approx V_{0}\left(\beta_{0}-\beta_{1}q^{\alpha-2}\right)
\end{equation}
 for $2<\alpha\le4$, where we have neglected higher order terms in
$q$. Here $\beta_{0}$ and $\beta_{1}$ are constants, and $V_{0}$
is defined in Eq.(\ref{alpha}). For $\alpha>4$, the leading order
term goes as $q^{2}$ (as can be seen, e.g., from the fact that for
$\alpha>4$ the second moment of the potential exists). We now
compute the potential energy

\begin{align}
U & =\frac{1}{2}\sum_{i,j}\left\langle V\left(\vec{r}_{i}-\vec{r}_{j}\right)\right\rangle \nonumber \\
 & =\sum_{k,k',q,\sigma,\sigma'}\frac{\tilde{V}\left(q\right)}{\Omega^{3}}\left\langle c_{k+q,\sigma}^{\dagger}c_{k,\sigma}c_{k'-q,\sigma'}^{\dagger}c_{k',\sigma'}\right\rangle ,
\end{align}
 where $\Omega$ is the volume of the system, and we have introduced
$c_{k,\sigma}$, the annihilation operator of an electron with
momentum $k$ and spin $\sigma$. The calculation can be simplified
significantly in the limit of small $mh/k_{0}^{2}$, in which the
Fermi surface becomes parametrically eccentric. In that limit,
only the dependence of $\tilde{V}_{q}$ on the momentum parallel to
the major (long) axis of the Fermi surface is important. The
result can be written as
$U=U_{\uparrow\uparrow}+U_{\uparrow\downarrow}$, where
$U_{\uparrow\uparrow}$ and $U_{\uparrow\downarrow}$ are the
interaction energies between same spins and opposite spins,
respectively, given (to leading order in $mh/k_{0}^{2}$) by

\begin{align}
U_{\uparrow\uparrow} & =A\left(\beta_{1}V_{0}\right)n\left[\left(\frac{k_{0}^{2}}{mh}\right)^{\frac{1}{4}}\sqrt{n}\right]^{\alpha-2},\nonumber \\
U_{\uparrow\downarrow} & =B\frac{\left(\beta_{0}V_{0}\right)n^{2}}{k_{0}^{2}}\sqrt{\frac{k_{0}^{2}}{mh}}.
\end{align}
 Here, $A$ and $B$ are dimensionless constants. We can now minimize
the total energy per particle, $\varepsilon=K+U_{\uparrow\uparrow}+U_{\uparrow\downarrow}$,
with respect to the variational parameter $h$. In the $n\rightarrow0$
limit, we find that $h^{\star}$, the optimal Zeeman field, is
\begin{equation}
h^{\star}\sim\begin{cases}
\frac{k_{0}^{2}}{m}\left(m\beta_{1}V_{0}\right)^{\frac{4}{\alpha}}n^{2-\frac{4}{\alpha}}, & \alpha<4\\
\beta_{0}V_{0}n, & \alpha>4.
\end{cases}\label{eq:hstar}
\end{equation}

 Inserting $h^{\star}$ back into the expression for the total energy,
we get that the ground state energy is
\begin{equation}
\varepsilon^{\star}\sim\begin{cases}
\frac{1}{m}\left(mV_{0}\right)^{\frac{2}{\alpha}}n^{2\left(1-\frac{1}{\alpha}\right)}, & \alpha<4\\
\frac{k_{0}^{2}}{m}\sqrt{m\beta_{0}V_{0}}\left(\frac{n}{k_{0}^{2}}\right)^{\frac{3}{2}}.
& \alpha>4.
\end{cases}\label{eq:E_FM}
\end{equation}

Finally, we argue that an out-of-plane Zeeman field, $h_z\ne 0$,
is not energetically favorable.
First, note that for an in-plane field, in the low-density limit ($n\rightarrow 0$) the kinetic energy per particle is unaffected %by the presence of the field.  This is so
because the Zeeman field becomes parallel to the spin-orbit field
for all occupied states, allowing each particle to gain the Zeeman
energy without changing its wavefunction. In contrast, in order
for an electron to align its spin with an out-of-plane Zeeman
field, its wavefunction must include hybridization with the
excited band (with energy $\sim k_0^2/m$ above the lower band). In
the hybrized state, the upper band is occupied with a probability
$\sim (mh_z/2k_0^2)^2$, leading to a kinetic energy cost
$(k_0^2/2m)\times(mh_z/2k_0^2)^2$. For contact interactions, the
potential energy per particle is given by $E_p = \frac14
(U_0/n)(n^2 - m_z^2)$, where $m_z \approx nh_z/(k_0/m^2)$ (here we
use that for small $h_z$, the spin of each electron obtains a
$z$-component given by the ratio of $h_z$ to the spin-orbit
field). Therefore the potential energy gain from $z$-polarization
is approximately $\frac14 U_0n(mh_z/k_0^2)^2$. Compared with the
kinetic energy cost, we see that the potential energy gain
includes an extra factor of $n$, indicating that the cost of
polarizing in the $z$-direction overwhelms the benefits in the
low-density limit.

% Finally, we note that an out-of-plane Zeeman field is not
% energetically favorable. This is because having a non-zero $z$
% component of the effective Zeeman field, $h_z\ne 0$, costs a
% finite kinetic energy per particle (of the order of $h_z^2$ in the
% small $h_z$ limit), since a Zeeman field in the $z$ direction
% mixes the two bands of the Rashba dispersion. The exchange energy
% gain, on the other hand, vanishes in the $n\rightarrow 0$ limit.
% Note that this is not so for an in-plane field: in that case the
% kinetic energy cost \emph{per particle} also vanishes in the
% $n\rightarrow 0$ limit (see Eq. \ref{eq:K}).

%, in the low-density limit, the system has to be nearly spin
%polarized to avoid a large potential energy cost. This requires a
%large Zeeman field, $h\gg k_{0}\sqrt{n}/2m$. But such a state has
%a large kinetic energy cost, of the order of $\frac{k_{0}^{2}}{m}$
%per particle, and therefore it cannot compete with a state with an
%in-plane polarization.

\bibliography{rashba_new1}

\begin{thebibliography}{35}
\bibitem[{com({\natexlab{a}})}]{comment-ell}
\bibinfo{note}{Note that this expression is not valid near $\eta=1$, and
  therefore it does not have an $\eta\rightarrow 1/\eta$ symmetry. For $\eta\ll
  1$, of course, a similar expression holds with $\eta\rightarrow 1/\eta$.}
\bibitem[{com({\natexlab{c}})}]{comment-lowest-subband}
\bibinfo{note}{For the optimal strip width, $L^\star_y\sim n^{-1/3}$, the Fermi
  energy is of the same order of the inter-subband splitting, so the assumption
  that only the lowest subband is occupied may not be justified. Nevertheless,
  one can construct a variational wavefunction with a ground state energy which
  scales as in Eq. \ref{eq:E_SM} by choosing $L_y\sim n^{-1/3}$ with a
  sufficeintly small prefactor such that only the lowest subband is occupied.}
\bibitem[{com({\natexlab{d}})}]{comment-current}
\bibinfo{note}{The fact that the current is zero can be seen, in the low
  density limit, from the fact that the Fermi surface is an ellipse enlosing
  the point $\vec{k}_0$ at which the group velocity is zero. See Appendix B.}
\bibitem[{com({\natexlab{e}})}]{comment-metallic-surface}
\bibinfo{note}{In the case of a metallic surface alloy, we are assuming that
  the effect of the bulk electrons is mostly to screen the Coulomb interaction
  on the surface.}
\end{thebibliography}

\end{document}